\documentclass[12pt]{scrartcl}
\usepackage{a4}
\usepackage{amsthm}
\usepackage{amsmath}
\usepackage{amssymb}
\usepackage{amsfonts}
\usepackage{mathrsfs}
\usepackage{latexsym}
\usepackage{color}
\usepackage{bbm,exscale}
\definecolor{Myblue}{rgb}{0,0,0.6}
\usepackage[a4paper,colorlinks,citecolor=Myblue,linkcolor=Myblue,urlcolor=Myblue,pdfpagemode=None]{hyperref}
\usepackage[square,numbers,sort&compress]{natbib}
\usepackage[all,cmtip]{xy}
\usepackage{tikz}

  \vfuzz \hfuzz
\def\1{\ifmmode\mathrm{1\!l}\else\mbox{\(\mathrm{1\!l}\)}\fi}

\newcommand{\be}{\begin{equation}}
\newcommand{\ee}{\end{equation}}
\newcommand{\bes}{\begin{equation*}}
\newcommand{\ees}{\end{equation*}}

\newcommand{\K}[3]{\left#1#3\right#2}

\newcommand{\s}{\hspace{2mm}}

\newcommand\arxiv[2]      {\href{http://arXiv.org/abs/#1}{#2}}
\newcommand\doi[2]        {\href{http://dx.doi.org/#1}{#2}}

\allowdisplaybreaks

\deffootnote[1em]{1em}{1em}
{\textsuperscript{\thefootnotemark}}

\theoremstyle{definition}
\newtheorem{definition}{Definition}

\newtheorem{theorem}[definition]{Theorem}

\numberwithin{equation}{section}
\numberwithin{definition}{section}
\numberwithin{figure}{section}

\newcommand\void[1]{}

\begin{document}

\title{On the Quantization of Special K\"ahler Manifolds}
\author{Michael M. Kay
\\[0.5cm]
  \normalsize{\tt \href{mailto:michael.kay@physik.uni-muenchen.de}{michael.kay@physik.uni-muenchen.de}}\\[0.1cm]
  {\normalsize\slshape Arnold Sommerfeld Center for Theoretical Physics, }\\[-0.1cm]
  {\normalsize\slshape LMU M\"unchen, Theresienstra\ss e~37, D-80333 M\"unchen}\\[-0.1cm]
  {\normalsize\slshape Excellence Cluster Universe, Boltzmannstra\ss e~2, D-85748 Garching}}

\date{}
\maketitle

\vspace{-11.8cm}
\hfill {\scriptsize LMU-ASC 34/13}

\vspace{12cm}

\begin{abstract}
Abstract: We show how affine and projective special K\"ahler manifolds emerge from the structure of quantization. We quantize them and construct natural (wavefunction) representations for the corresponding coherent states. These in turn are shown to satisfy the precise generalizations of the \emph{BCOV holomorphic anomaly equation} (hep-th/9309140), thus extending the work in hep-th/9306122. As a byproduct of the analysis we construct the explicit general solution to the holomorphic anomaly equation.
\end{abstract}

\newpage

\tableofcontents

\section{Introduction}
Special K\"ahler manifolds have been studied in the physics literature since the seminal papers \cite{sierra-townsend}, \cite{specgeopaper1}, \cite{dewit-vanproeyen}, \cite{gates}. These split into two main categories. The first are known as affine special K\"ahler manifolds (or rigid special K\"ahler manifolds) while the second as projective special K\"ahler manifolds (or local special K\"ahler manifolds). While the former arise as moduli spaces of vector-multiplets in rigid $N=2$ supersymmetric four dimensional gauge theories  (e.g. \cite{sierra-townsend}, \cite{seiberg-witten1}, \cite{seiberg-witten2}), the latter arise analogously in the corresponding (locally supersymmetric) supergravity theories (e.g. \cite{specgeopaper1}\cite{dewit-vanproeyen}, \cite{stromingerspecgeo}). The structure of projective special K\"ahler manifolds has also been rediscovered within the framework of string theory (\cite{cecottiferraragirardello}, \cite{cecottivafa1}, \cite{bcov9309140}), thus providing a microscopical description of the vector-multiplet moduli spaces in the aforementioned four dimensional theories. Mathematically, special K\"ahler manifolds have been defined both extrinsically (\cite{cortes}, \cite{bauescortes}) and intrinsically \cite{freed}. Of great relevance is also \cite{costelloli}, where central results of \cite{bcov9309140} have been understood and extended within a rigorous mathematical framework.

\medskip
\noindent
Focusing on the string theory perspective, projective special K\"ahler manifolds arise as moduli spaces of certain two dimensional closed topological string theories whose underlying topological field theories are obtained as topological twists of $N=(2,2)$ two dimensional conformal field theories, most notably sigma models into complex three dimensional Calabi-Yau manifolds. The fundamental object of study in these topological string theories is the generating functional of all scattering amplitudes of closed topological string states. In fact, much of the attention is reserved to states that correspond to so-called marginal fields. These are fields that induce infinitesimal deformations of the corresponding topological field theory to a ``neighboring one". In \cite{bcov9309140} it was shown that this restricted generating functional satisfies a differential equation, which was named the \emph{holomorphic anomaly equation}. More precisely the restricted generating functional is of the form:
\begin{equation*}
Z(u, p),
\end{equation*}
where $u$ stands for a choice of marginal field, while $p$ labels the topological conformal field theory within which the scattering of marginal fields takes place. And the name ``holomorphic anomaly" is due to the fact that if one departs from genus zero scattering surfaces one finds that the generating functional has non-holomorphic dependence on the space of $p$'s, namely the moduli space of  topological field theories attached to a given initial one. In \cite{witten} it was shown how a similar structure to that of the holomorphic anomaly equation arises if one performs a version of geometric quantization of the moduli space of $B$-model topological conformal field theories, where the special K\"ahler manifold is a moduli space of complex structures of complex three dimensional Calabi-Yau manifolds.

\medskip
\noindent
In this paper we expand on the work of \cite{witten}, showing, in fact, how the very structure of special K\"ahler manifold, both affine and projective, arises as the simplest quantizable geometry. Moreover, in our framework we recover the precise form of the holomorphic anomaly equation of \cite{bcov9309140}, while at the same time providing its general solution. It is important to stress that just as in \cite{witten}, also in the present work, the perspective is not a microscopical one. In particular, our derivation of the holomorphic anomaly equation is not ascribed to the detailed knowledge of the moduli spaces of Riemann surfaces. Crucial to our construction is instead the approach of \cite{fedosov94} to the quantization of symplectic manifolds. In close analogy to \cite{witten}, though in much greater generality, in this framework the holomorphic anomaly equation translates to a parallel transport equation with respect to a flat connection $A$ on a Hilbert bundle over the classical phase space.

\medskip
\noindent
The present work develops as follows. In section \ref{quantizationreview} we recall the very basics of quantization. In section \ref{quantsymplectic} we intend to provide a simple review of the approach in \cite{fedosov94} to the quantization of symplectic manifolds. In section \ref{firstorderholo} we show how affine special K\"ahler manifolds arise as the ``simplest" quantizable spaces. Complementary to the developments of the aforementioned sections is the representation, in \ref{CTB}, of the coherent states corresponding to the affine special K\"ahler manifolds. The crucial notion here is that of \emph{coherent tangent bundle}. Subsequently in \ref{MEsection} we show how the thus constructed wavefunction, at this point only for strictly Riemannian affine special K\"ahler manifolds, satisfies a version of the holomorphic anomaly equation, which we shall call throughout simply \emph{master equation}. In \ref{sol1} and \ref{sol2} we then provide its general solution. In \ref{conicsk} we start the study of the projective special K\"ahler manifolds of interest in the physics literature. These arise as quotients of affine \emph{conic}, but Lorentzian, affine special K\"ahler manifolds. Thus first we unravel the structure of conic special K\"ahler manifold in a way best suited for our formalism, also providing a possible quantum interpretation \ref{quantumconic} for the conic structure. Subsequently we define the quantization of Lorentzian affine special K\"ahler manifolds (sections \ref{LCTB} and \ref{lorentzquantum}) and finally we show how to quantize the quotient, projective special K\"ahler manifold in \ref{projsk}. In this way we arrive at the desired master equation, while simultaneously having provided its general solution.
\section{Quantization review}\label{quantizationreview}
In this section we will review the general principles underlying quantization of classical phase spaces. The basic ingredients are a classical phase space, which we will assume to be a smooth manifold $M$, a space of quantum states, which by definition is complex projective space $\mathbb{P}^n$ of a priori arbitrary dimension $n$, and a quantization map. This is a map
\begin{equation*}
\phi: M \rightarrow \mathbb{P}^n
\end{equation*}
identifying the classical state space (or a portion thereof) as a subset of the space of quantum states\footnote{As classical mechanics should in principle be recovered from quantum mechanics, $\phi$ should be in some sense faithful, e.g. an immersion or even embedding.}. The image of this map, $\phi(M)$, is known as the space of coherent states. Part of the problem of quantization is the classification of such triples. Complementary to that, is the task of transporting the basic invariants of $\mathbb{P}^n$ via $\phi$ to $M$. The basic algebraic invariant of interest is the maximal compact subgroup $G$ of $\mathrm{Aut}(\mathbb{P}^n)$. This can also be viewed as the group of automorphisms of $\mathbb{P}^n$ endowed with the pairing:
\begin{equation*}
(x, y) := \frac{|\langle x,  y \rangle|}{||x||\,||y||},
\end{equation*}
where $\langle \cdot, \cdot \rangle$ denotes a sesquilinear product on $\mathbb{C}^{n+1}$. It is the result of Wigner's Theorem, that $G$ is composed exactly of the unitary and antiunitary transformations of $(\mathbb{C}^{n+1}, \langle \cdot, \cdot \rangle$). The \textit{space of quantum observables} is the Lie algebra $\mathfrak{g} = \mathrm{Lie}(G)$, while what is known as the \textit{algebra of quantum observables} is its universal enveloping algebra $U(\mathfrak{g})$. 

\medskip
\noindent
The algebra of classical observables is recovered as follows. Let $F \in U(\mathfrak{g})$, then its classical counterpart is a complex valued function $f: M \rightarrow \mathbb{C}$ given by:
\begin{equation*}
f(p) = \langle \phi(p),  F  \phi(p) \rangle.
\end{equation*}
The universal enveloping algebra is represented by the so-called star-product $\star$, which by definition must satisfy:
\begin{equation*}
(f \star g)(p) := \langle \phi(p), F \cdot G  \,  \phi(p) \rangle.
\end{equation*}
In the following we will suppress one degree of arbitrariness in the choice of the quantization triple, namely the dimension of projective space. In fact, without loss of generality, we are allowed to consider the direct limit:
\begin{equation*}
\mathbb{P}^{\infty} := \lim_{n \rightarrow \infty} \mathbb{P}^n.
\end{equation*}
This in turn can be viewed as $\mathcal{H}/\mathbb{C}^*$, where $\mathcal{H}$ is an infinite dimensional separable Hilbert-space. It is important to remark, as it will be crucial in what follows, that any two such Hilbert-spaces are isomorphic. Before venturing into the more general case, it will be useful to recall the very well known quantization of $\mathbb{R}^{2d}$.

\subsection{The simplest case: $M = \mathbb{R}^{2d}$}\label{simplestcase}
Given the definition of quantization above, a priori there are a multitude of quantization maps $\phi$ of $\mathbb{R}^{2d}$. However its canonical quantization presupposes a much more rigid structure than that of a smooth manifold. Indeed $\mathbb{R}^{2d}$ is identified with its group of translations $\Gamma$, or more precisely with an orbit, e.g. $\Gamma \cdot e$, where $e$ denotes the identity element $e = 0 \in \mathbb{R}^{2d}$. Then the quantization maps reduce to the projective representations:
\begin{equation*}
\rho : \Gamma \rightarrow \mathrm{Aut}(\mathbb{P}^{\infty}).
\end{equation*}
As is well known these are in one-to-one correspondence with the family of linear representations:
\begin{equation*}
\hat{\rho} : \hat{\Gamma} \rightarrow \mathrm{Aut}(\mathcal{H})
\end{equation*}
labeled by a central extension $\hat{\Gamma}$ of $\Gamma$. These in turn are fully specified by the choice of a skew-symmetric bilinear form $\omega^{-1} \in \bigwedge^2 (\mathrm{Lie}(\Gamma))^*$. Passing to the Lie algebra description altogether, $\mathrm{Lie}(\hat{\Gamma})$ is then specified by the following commutation relations:
\begin{equation*}
[\hat{x}^i, \hat{x}^j] = i\omega^{-1}(x^i, x^j).
\end{equation*}
By a slight abuse of notation, we have multiplied the generators by $i = \sqrt{-1}$, so that these will be represented as self-adjoint operators. We will restrict attention to the case where $\omega^{-1}$ is non-degenerate. Although this is no real loss of generality in the present case, it will be in the following sections, where the space $(\mathbb{R}^{2d}, \omega^{-1})$ is generalized to a Poisson manifold, while we will be solely interested in the symplectic case\footnote{Roughly speaking the requirement that the phase-space be symplectic replaces the notion that the quantization map $\phi$ should be `` faithful".}. In the non-degenerate case, $\hat{\Gamma}$ is known as a \textit{Heisenberg group}, and these are in fact all equivalent. This simply follows from the fact that any non degenerate skew-symmetric matrix can be brought to canonical form $\epsilon$ by an invertible matrix $\Lambda$, as
\begin{equation*}
\Lambda^T \omega \Lambda = \epsilon.
\end{equation*}
It will be useful in the following to introduce further canonical objects: $\eta$, the standard euclidean metric, and the complex structure $I$ given by:
\begin{equation*}
\eta = I\epsilon.
\end{equation*}
Part of the Stone-von-Neumann-Mackey Theorem states that $\hat{\Gamma}$ has a unique, up to isometry, unitary irreducible and infinite dimensional representation on a separable Hilbert-space. In fact, since infinite dimensional separable Hilbert-spaces are all equivalent, we can view each such $\mathcal{H}$ as furnishing such an irreducible representation. Indeed this is realized as follows. First, presupposing canonical form, split $\mathrm{Lie}_{\mathbb{C}}(\hat{\Gamma})$ into raising and lowering subalgebra spanned by the operators:
\begin{equation*}
\hat{x}^i(\eta + i\epsilon)_{ij} \s\s\s \textrm{and}\s\s\s \hat{x}^i(\eta - i\epsilon)_{ij}
\end{equation*}
respectively. The former are commonly known as annihilation while the latter as creation operators. Then choose an orthonormal basis $\{| n \rangle\}$ of $\mathcal{H}$ enumerated by $n \in \mathbb{N}_0^{d}$. Declare $|0 \rangle$ to be the highest weight vector and let the action of $\hat{x}^i(\eta - i\epsilon)_{ij}$ be specified by:
\begin{equation*}
\hat{x}^i(\eta - i\epsilon)_{ij} |n \rangle \sim |n + e_j \rangle,
\end{equation*}
where $e_j$ denotes the unit vector in the $j$th direction and $\sim$ indicates equal up to a suitable unique proportionality factor. At this point we can turn to the representation $\hat{\rho}$:
\begin{equation*}
\hat{\rho}(p) = \exp(i\omega_{ij}x^i \hat{x}^j),
\end{equation*}
where $p = (x^1, \dots, x^{2d})$. Given this representation, it is straightforward to obtain the quantization map:
\begin{equation}
\phi(p) = \hat{\rho}(p) |\psi \rangle,
\label{cohstate1}
\end{equation}
where $|\psi \rangle$ is an arbitrary, nonzero, state in $\mathcal{H}$. This choice is irrelevant, it can be removed by an automophism of $\mathbb{P}^n$. If we choose $|\psi\rangle = |0 \rangle$, we recover the canonical notion of coherent state:
\begin{equation}
|x\rangle = \exp(i\omega_{ij}x^i \hat{x}^j)|0\rangle \label{poisscohstate}.
\end{equation}
This in particular satisfies:
\begin{equation*}
\hat{x}^i(\eta + i\epsilon)_{ij} |x\rangle = x^i(\eta + i\epsilon)_{ij} |x\rangle,
\end{equation*}
namely, it is an eigenstate of the annihilation operators.

\medskip
\noindent
For the sake of completeness we will sketch how to extract the star-product in the particular case $M = \mathbb{R}^2$ \footnote{The more general case $M=\mathbb{R}^{2d}$ is then obtained in a straightforward fashion.}. While the above notation is convenient as a reference for future sections, we will, solely for this independent appendix to this review section, use the common notation with annihilation operator $a$ and creation operator $a^{\dagger}$, which up to a factor are equivalent to the ones defined above. Then the canonical coherent states are usually denoted as $|\alpha \rangle$, where $\alpha = (1/\sqrt{2})(y^1 + iy^2)$ and $(y^1,y^2)=x$ are the coordinates of a point $p \in \mathbb{R}^2$. Then, from the defining property:
\begin{equation*}
a |\alpha \rangle = \alpha |\alpha\rangle,
\end{equation*}
one can recover the state $|\alpha \rangle$ as:
\begin{equation*}
|\alpha \rangle = \exp\K{(}{)}{-\frac{1}{2}|\alpha|^2}\sum_{n \geq 0} \frac{\alpha^n}{\sqrt{n!}}|n \rangle,
\end{equation*}
where we have normalized $|\alpha \rangle$ to 1. Moreover, in the usual notation:
\begin{equation*}
\hat{\rho}(p) =: U(\alpha) = \exp(\alpha a^{\dagger} + \overline{\alpha}a).
\end{equation*}
Then
\begin{align*}
(f \star g)(p) &= \langle \alpha | F \cdot G | \alpha \rangle\\
&= \sum_{n \geq 0} \langle 0 | U(-\alpha) F  U(\alpha) \; |n \rangle \langle n | \; U(-\alpha) G U(\alpha) | 0 \rangle\\
&= \sum_{n \geq 0} \frac{1}{n!}\langle 0 | U(-\alpha) F  U(\alpha) \; (a^{\dagger})^n | 0 \rangle \langle 0 | a^n \; U(-\alpha) G U(\alpha) | 0 \rangle\\
&= \sum_{n \geq 0} \frac{1}{n!}\langle 0 | \mathrm{ad}_{-a^{\dagger}}^n[U(-\alpha) F U(\alpha)] | 0 \rangle \langle 0 | \mathrm{ad}_{a}^n[U(-\alpha) F U(\alpha)] | 0 \rangle\\
&= \sum_{n \geq 0} \frac{1}{n!} \K{(}{)}{\frac{\partial^n}{\partial \alpha^n}f}(p)\K{(}{)}{\frac{\partial^n}{\partial \overline{\alpha}^n}g}(p)\\
&= \K{(}{)}{f \exp\K{(}{)}{\frac{1}{2}\langle \stackrel \leftarrow \nabla, (\eta^{-1} + i \epsilon^{-1}) \stackrel \rightarrow \nabla\rangle} g}(p).
\end{align*}
In order to appreciate the significance of the star product in physics, one should introduce Planck's constant $\hbar$, which we have implicitly set to $1$. The latter is reintroduced precisely by the following change of coordinates: $(y_1, y_2) \mapsto (\sqrt{\hbar}y_1, \sqrt{\hbar}y_2)$. Then the star product reads:
\begin{equation*}
(f \star g)(p) = \K{(}{)}{f \exp\K{(}{)}{\frac{\hbar}{2}\langle \stackrel \leftarrow \nabla, (\eta^{-1} + i \epsilon^{-1}) \stackrel \rightarrow \nabla\rangle} g}(p).
\end{equation*}
Using the above one can now, in particular, recover Hamilton's equations of classical mechanics as the classical limit ($\hbar \rightarrow 0$) of Heisenberg's equations.
\section{A more general case: quantization of symplectic manifolds}\label{quantsymplectic}
We are now faced with the problem of generalizing this beautiful yet very special construction for $\mathbb{R}^{2d}$ to the general case of a symplectic manifold $M$ of dimension $2d$. For this, we follow Fedosov's method \cite{fedosov91, fedosov94}. Accordingly we construct quantization maps as follows. First we choose a point $p \in M$ and declare that this be mapped to the point $\phi(p) =: |p \rangle \in \mathbb{P}^{\infty}$. Next, we declare that any other point $p'$ in the vicinity (to be explained later) of $p$ be mapped to the point:
\begin{equation*}
p' \rightarrow \phi(p') =: U(p', p) | p \rangle,
\end{equation*}
for some $U(p',p) \in \mathrm{Aut}(\mathbb{P}^{\infty})$. In particular it must be continuously connected to the identity, therefore $U(p', p)$ is unitary. In fact, thanks to the $QR$ decomposition of matrices, this is no loss of generality. Let's now erect a (at this point arbitrary) coordinate system $\{x^k\}$ in a neighborhood $V_{p}$ of $p$. And let's define the object
\begin{equation*}
-A(p) = \K{.}{|}{\frac{\partial}{\partial x^k} U(p', p)}_{p' = p} dx^k.
\end{equation*}
We can interpret $A$ as a flat connection on a $\mathbb{P}^{\infty}$-bundle over $X$. For computational purposes however, it is more convenient to work on the corresponding $\mathcal{H}$-bundle where, by a slight abuse of notation, the connection $A$ is allowed to have holonomies in the centre of $\mathrm{Aut}(\mathbb{P}^{\infty})$ namely $\mathbb{C}^*$. That is, $A$ satisfies the Maurer-Cartan equation:
\begin{equation*}
dA + A\wedge A \in \Omega^2(M, \mathbb{C}).
\label{MC}
\end{equation*}
In other words, $A$ is projectively flat as a connection on the $\mathcal{H}$-bundle. Without loss of generality we can however assume that, as a connection on the $\mathcal{H}$-bundle, $A$ is flat, namely:
\begin{equation}
dA + A\wedge A = 0.
\label{MC2}
\end{equation}
For this we simply have to twist the $\mathcal{H}$-bundle by a hermitian line-bundle with a connection whose curvature precisely cancels that of $A$. We shall hitherto refer to the state $|p \rangle$ parallel transported by $A$, viewed as an element in $\mathcal{H}$, as $|p \rangle_A$. Clearly $U(p', p)$ can be written in the form:
\begin{equation*}
U(p', p) = \mathcal{P}\exp\K{(}{)}{-\int_{\gamma}A},
\end{equation*}
where $\mathcal{P}$ stands for path-ordered, $\gamma: [0,1] \rightarrow M$ is a path with endpoints $\gamma(0) = p$, $\gamma(1) = p'$, and since $A$ is flat the result of the integration only depends on the homotopy class $[\gamma]$. In particular, if we restrict attention to a simply connected, or even better, contractible neighborhood of $p$, then the result of integration is completely independent of the chosen path and in that case we are allowed to refer to the integral as $\int_p^{p'}$.  One could also do this globally if one replaces $M$ with its universal cover altogether.

\medskip
\noindent
So far the discussion was very general, in that we have not required any special properties of $M$ other than it be smooth and we have traded the notion of quantization map for that of a flat connection on an $\mathcal{H}$-bundle. The interesting step is now to find a good classification of the solutions to (\ref{MC2}).  We will assume at this point that $M$ is symplectic, and we choose $V_p$ to be a Darboux patch, namely a coordinate neighborhood where the symplectic form $\omega$ is flat. Attached to this flat symplectic form we have a corresponding Heisenberg algebra with generators $\hat{x}^i$ and $\mathcal{H}$ is the corresponding irreducible representation. The intuition behind this is to envisage the tangent space $T_pM$ at each point $p \in M$ as a copy of $\mathrm{Lie}(\Gamma)$. In a suitable sense every self-adjoint operator of $\mathcal{H}$ is an element of $U(\mathrm{Lie}(\Gamma))$. In informal terms, this follows from the following decomposition of the projector on the highest weight state $|0 \rangle$:
\begin{equation*}
|0 \rangle \langle 0 | = \sum_{\vec{k} \in \mathbb{N}_0^d} (-1)^k\frac{(a^{\dagger})^ka^k}{k!} = \sum_{\vec{k} \in \mathbb{N}_0^d} (-1)^k {a^{\dagger}a \choose k}.
\end{equation*}
In the above we have used multi-index notation. This decomposition allows us to expand $A$ as follows:
\begin{equation}
A(p) = i\sum_{l = 0}^{\infty} \sum_{i_1 \leq \dots \leq i_l}(\alpha_{i_1, \dots, i_l, k}(p)\hat{x}^{i_1} \cdots \hat{x}^{i_l} \; + h.c) dx^k.
\label{pertexp}
\end{equation}
Equation (\ref{MC2}) thus decomposes into an infinite number of equations. More precisely (\ref{pertexp}) is well defined in the topology defined by the seminorms:
\begin{equation*}
|\langle \psi_1| \, \cdot  \, | \psi_2 \rangle| \s \textrm{with } \psi_1 \in \mathcal{S}_d, \psi_2 \in \mathcal{H},
\end{equation*}
where by $\mathcal{S}_d \subset \mathcal{H}$ we denote the space of states whose coefficients $c_n$ in the expansion $\psi_1 = \sum_n c_n |n\rangle$ tend to zero as $||n|| \rightarrow \infty$, faster than any polynomial of $n \in \mathbb{N}_0^d$   \footnote{This space is also known as the space of rapidly decreasing sequences, which can be equipped with a Fr\'echet topology.}. Thus the notation $\mathcal{S}_d$ is suggestive for Schwarz-space, although this should not be taken literally. At this point we remark that the canonical, Poissonian coherent states of $M = \mathbb{R}^{2d}$ are elements of $\mathcal{S}_d$. What this restriction on the topology implies, is that for the perturbative ansatz (\ref{pertexp}) to be well-defined, we should represent the state $|p \rangle_A$ as a wavefunction whose corresponding complete sequence of linear functionals has as corresponding sequence of states, elements of $\mathcal{S}_d$. We will define such wavefunctions in section \ref{CTB}.

\medskip
\noindent
Returning to the infinite sequence of equations encoded in (\ref{MC2}), we will see in the following that each equation specifies a certain geometrical structure on $M$. The philosophical perspective one could take about the above expansion is that, as we increase the order in perturbation theory we are chiseling step by step, through equation (\ref{MC2}), the geometry of a Darboux patch of $M$. In particular we assume that at each step in perturbation theory the Darboux patch be smooth, however this does not impose that the limiting structure be. In other words, our initial assumption that $M$ be smooth could in principle be omitted for the limiting geometries. As a check, and for matters of convention, let's recover the simplest case $M = \mathbb{R}^{2d}$ in this formalism. There the perturbation expansion stops at first order:
\begin{equation*}
A = i (\alpha_k + \omega_{ik}\hat{x}^i)dx^k.
\end{equation*}
Thus, solving (\ref{MC2}) yields the following two equations:
\begin{align*}
d\alpha &= -\frac{1}{2}\omega\\
\partial_k (\omega_{il}\hat{x}^i) dx^k \wedge dx^l &= 0.
\end{align*}
The second equation is automatically satisfied. Thus, apart from an irrelevant phase:
\begin{equation*}
U(p', p) = \mathcal{P}\exp\K{(}{)}{-i\int_p^{p'}\K{(}{)}{-\frac{1}{2}\omega_{ik}x^i + \omega_{ik}\hat{x}^i}dx^k}.
\end{equation*}

\section{Somewhere in between: special geometries}\label{secondorder}
In this section we shall investigate the geometry of phase-spaces whose associated connection $A$ stops at second order in the perturbative expansion (\ref{pertexp}). We will show in section \ref{gauges} that these spaces are actually equivalent to the ones whose connection stops at first order. As will become clear in the following sections this class includes affine (Riemannian) special K\"ahler manifolds.

\medskip
\noindent
First we will investigate equation (\ref{MC2}) to second order. Furthermore we will assume that, to first order, $A$ reduces to the flat case. The connection $A$ then takes the form:
\begin{equation}
A = i(\alpha_k + \omega_{ik}\hat{x}^i + D_{ijk}\hat{x}^i \hat{x}^j)dx^k,\label{Asecondorder}
\end{equation}
where, given that $A$ is hermitian, and without loss of generality, $D_{ijk} \in \mathbb{R}$ and $D_{ijk} = D_{jik}$. Equation (\ref{MC2}) becomes:
\begin{align}
d\alpha &= -\frac{1}{2}\omega \nonumber\\
\label{GM1}
D_{krl} - D_{lrk} &= 0\\
\label{GM2}
\partial_kD_{ijl} - \partial_lD_{ijk} - 2(D_{isk}D_{rjl} + D_{jsk}D_{ril})\omega^{sr} &= 0.
\end{align}
We now introduce the following object:
\begin{equation*}
G_{ki}^j = 2D_{ilk}\omega^{lj}.
\end{equation*}
The symmetry of $D_{ijk}$ in its first two indices translates to:
\begin{equation*}
G_{ki}^l\omega_{lj} - G_{kj}^l\omega_{li} = 0,
\end{equation*}
that is:
\begin{equation*}
G_k \omega + \omega G_k^T = 0.
\end{equation*}
In other words, $G_k$ is a symplectic matrix. In terms of $G_k$, equations (\ref{GM1}) and (\ref{GM2}) read:
\begin{align*}
G_{lk}^i - G_{kl}^i &= 0\\
\partial_k G_l  - \partial_l G_k - [G_k, G_l] &= 0.
\end{align*}
We will learn in section \ref{gauges} that $G$ is a connection on the tangent bundle of $M$. Then
the first equation is the statement that $G$ is torsion-free, while the second means that $G$ is flat. So to summarize, the second order quantizations correspond to symplectic manifolds with a flat symplectic connection \footnote{Recall that a connection that is both compatible with the symplectic form and torsion free is known as a symplectic connection, while if it is not torsion free it is called quasi-symplectic.}.

\subsection{K\"ahler manifolds: holomorphic connections}\label{firstorderholo}
We now sharpen our analysis to the case where the phase-space $M$ is a complex symplectic manifold, that is, a K\"ahler manifold, when endowed with the appropriate compatible metric $g = J\omega$, where $J \in \Gamma(M, \mathrm{End}(TM))$ denotes its complex structure. We then ask when it is that the above constructed quantization map is compatible with $J$. We define compatibility as follows.

\begin{definition}\label{compatibholo}
A quantization map $\phi: M \rightarrow \mathbb{P}^{\infty}$ defined by a projectively flat unitary connection $A$, is compatible with the complex structure of $M$, if $A$ admits the following decomposition:
\begin{equation*}
A = \frac{1}{2}(B + B^{\dagger}),
\end{equation*}
where $B$ is a holomorphic, projectively flat connection. That is, $\phi$ induces a holomorphic map to $\mathbb{P}^{\infty}$.
\end{definition}
\noindent
It is straightforward to check that the above decomposition for $A$ is unique with $B$ given by:
\begin{equation*}
B = A_r(\delta_k^r + iJ_k^r)dx^k.
\end{equation*}
We will now check, at first order, what conditions on the geometry of $M$ must be imposed in order for the compatibility condition of definition \ref{compatibholo} to be fulfilled. The holomorphic connection is given by:
\begin{align*}
B &= i\K{(}{)}{\alpha_r + \omega_{ir}\hat{x}^i}(\delta^r_k + iJ^r_k)dx^k\\
&= i\K{(}{)}{(\delta^r_k + iJ^r_k)\alpha_r - (\omega + ig)_{ki}\hat{x}^i}dx^k.
\end{align*}
Let $\Omega$ denote the curvature of $B$, then the projective flatness condition reads:
\begin{align}
\K{(}{)}{i\partial_k(\alpha_l + iJ_l^r \alpha_r) - \frac{1}{2}[(g + i\omega)_{ki}\hat{x}^i, (g + i\omega)_{lj}\hat{x}^j]}dx^k \wedge dx^l &= \Omega \label{holocurvature}\\
\partial_k(\omega - ig)_{il} - \partial_l(\omega - ig)_{ik} &= 0. \nonumber
\end{align}
The second equation reduces to
\begin{equation*}
\partial_k g_{il} - \partial_l g_{ik} = 0.
\end{equation*}
That is, there are Darboux coordinates where:
\begin{equation*}
g_{il} = \partial_l f_i = \partial_i f_l = \partial_i \partial_l K,
\end{equation*}
where $f_i$ and $K$ are real valued functions on $M$. In fact it is straightforward to observe that $K$ is a K\"ahler potential for $M$. Moreover, while an arbitrary K\"ahler potential is defined up to holomorphic functions, $K$ is defined only up to linear ones.

\medskip
\noindent
To end the above analysis we shall return to equation (\ref{holocurvature}), which now reduces to:
\begin{equation*}
\Omega = -\frac{i}{2}\omega - J_l^r\partial_r\alpha_k dx^k \wedge dx^l,
\end{equation*}
which, in the gauge:
\begin{equation}
\alpha = -\frac{1}{2}J_k^l \partial_l K dx^k
\label{canonalpha}
\end{equation}
becomes:
\begin{equation*}
\Omega = -\frac{i}{2}\omega.
\end{equation*}
We shall denote this gauge for $\alpha$ as canonical. We now assume that a second order quantizable manifold admits a gauge in which $B$ is of the particular form just considered. This is the case exactly when, in the above Darboux coordinates the flat symplectic connection $G$ vanishes. That is the Darboux coordinates are $(d + G)$-flat. Then, in arbitrary coordinates the constraint on the metric reads:
\begin{align*}
d_{d + G}J = 0.
\end{align*}
We thus obtain exactly the definition of affine special K\"ahler manifold (see e.g. \cite{freed}). In the following section we shall finally show the equivalence of first and second order quantizable spaces. To conclude this section we shall formalize our findings with the following
\begin{theorem}
A K\"ahler manifold with quantization map $\phi : M \rightarrow \mathbb{P}^{\infty}$ whose corresponding flat connection $A$ is first-order in a suitable coordinate system, and is compatible with the complex structure of $M$, is precisely an affine special K\"ahler manifold.
\end{theorem}
\subsection{Symplectomorphisms as gauge transformations}\label{gauges}
Here we will show how symplectomorphisms act on the coherent state $|p\rangle$, thus allowing us in particular to transform the flat connection $A$ from special to arbitrary Darboux coordinates.  In particular we will show that every second order connection $A$ of the form (\ref{Asecondorder}) can be brought to first order, under a suitable symplectic change of coordinates. From now on, we shall denote by $A_s$ the first order connection $A$ in special Darboux coordinates. Let $\sigma : M \rightarrow M$ denote a local symplectomorphism on $M$, and let $\Sigma : TM \rightarrow TM$ denote its differential. Then $\sigma$ acts on $\mathcal{H}$ via the unitary map:
\begin{equation}
S := \exp (-if -\frac{i}{2}\K{(}{)}{\log(\Sigma) \omega}_{ij} \hat{x}^i\hat{x}^j),\label{SofSigma}
\end{equation}
where $f$ is an arbitrary real function and the second term is antihermitian if and only if $\sigma$ is a symplectomorphism. The function $f$ can be included as $\sigma$ should only act projectively on $\mathcal{H}$. To verify that $\sigma$ acts via $S$ we simply need to use the fact that $\mathcal{H}$ is an irreducible representation of the Heisenberg algebra, thus reducing the problem to the following single check:
\begin{equation*}
S \hat{x}^k S^{-1} = \Sigma_l^k \hat{x}^l.
\end{equation*}
For this we shall consider the one parameter family of symplectomorphisms defined by $\Sigma_t:= \exp(t \log(\Sigma))$ and will show that it is in correspondence with $S_t = \exp(-t(if + \frac{i}{2}\K{(}{)}{\log(\Sigma) \omega}_{ij} \hat{x}^i\hat{x}^j))$. To this aim we only need to verify that the two families agree in the immediate neighborhood of $t=0$:
\begin{align*}
\frac{d}{dt} \K{(}{)|}{S_t \hat{x}^k S_t^{-1}}_{t = 0} &= -\frac{i}{2} (\log(\Sigma)\omega)_{ij} [\hat{x}^i \hat{x}^j, \hat{x}^k]\\
&= -\frac{i}{2} (\log(\Sigma)\omega)_{ij}(i\omega^{ik} \hat{x}^j + i\omega^{jk} \hat{x}^i)\\
&= \frac{1}{2}(\log(\Sigma)^T-\omega^{-1}\log(\Sigma)\omega)\hat{x}\\
&= \log(\Sigma)_l^k \hat{x}^l,
\end{align*}
where, in the last step, we have used the fact that $\Sigma$ is a symplectomorphism. On the flat connection $A$, $S$ acts as a gauge transformation. Let's take $A = A_s$, then, under a coordinate transformation:
\begin{align}
(A_s)_k \mapsto \Sigma_k^l\K{(}{)}{S (A_s)_l S^{-1} + S \partial_l S^{-1}}.\label{gaugetrans}
\end{align}
In order to compute $S\partial_k S^{-1}$ we resort once again to the flows $\Sigma_t$ and $S_t$ and compare the time derivatives at arbitrary time $t$:
\begin{align*}
\frac{d}{dt} S_t \partial_k S_t^{-1} &= S_t \partial_k\K{(}{)}{if + \frac{i}{2}\K{(}{)}{\log(\Sigma)\omega}_{ij}\hat{x}^i\hat{x}^j}S_t^{-1}\\
&= i\partial_k f + \frac{i}{2} (\partial_k(\log(\Sigma)) \omega)_{ij} (\Sigma_t)_r^i (\Sigma_t)_s^j \hat{x}^r \hat{x}^s\\
&= i\partial_k f + \frac{i}{2} (\Sigma_t \partial_k(\log(\Sigma)) \omega \Sigma_t^T)_{rs}\hat{x}^r\hat{x}^s\\
&= i\partial_k f + \frac{i}{2} (\Sigma_t \partial_k(\log(\Sigma))\Sigma_t^{-1}\omega)_{rs}\hat{x}^r\hat{x}^s\\
&= \frac{d}{dt}\K{(}{)}{it \partial_k f - \frac{i}{2}(\Sigma_t (\partial_k \Sigma_t^{-1})\omega)_{rs}\hat{x}^r \hat{x}^s}.
\end{align*}
Therefore:
\begin{align}
(A_s)_k &\mapsto iS(\Sigma_k^l \alpha_l + \omega_{il}\Sigma_k^l\hat{x}^i)S^{-1} + \frac{i}{2}\Sigma_k^l\partial_l f - \frac{i}{2}\Sigma_k^l(\Sigma(\partial_l \Sigma^{-1})\omega)_{ij}\hat{x}^i\hat{x}^j \nonumber\\
&= i \K{(}{)}{\Sigma_k^l(\alpha_l  + \partial_l f) + \omega_{ik} \hat{x}^i - \frac{1}{2}\Sigma_k^l(\Sigma(\partial_l \Sigma^{-1})\omega)_{ij}\hat{x}^i\hat{x}^j}\label{AstoA}.
\end{align}
We thus recovered the general form (\ref{Asecondorder}) and verified that indeed $G$ is a connection on the tangent bundle to $M$. Moreover, in (\ref{AstoA}) we also observe that $i\alpha$ should be viewed as a connection on a line-bundle over $M$. More precisely, the transformation properties of $\alpha$ under the gauge transformation $\exp(-if)$ show that this line bundle is precisely the pullback of the unitary tautological bundle $\mathcal{H} \rightarrow \mathbb{P}^{\infty}$, or Hopf-fibration, via the quantization map. Henceforth we shall denote the operator $S$ corresponding to the differential $\Sigma$ as $S_{\Sigma}$.

\subsection{The ``coherent" tangent bundle}\label{CTB}
In this section we will give an explicit realization of the state $|p\rangle_A$ as a wavefunction $Z_A(u,p)$. The following discussion in fact applies to any K\"ahler manifold. We wish the wavefunction to correspond to a covariant tensor on $M$ thus allowing us to speak of the state $|p \rangle_A$ as a coordinate independent object. We thus define the wavefunction as follows:
\begin{equation*}
Z_A(u, p) = \, _{p, A} \!\,\langle u | p \rangle_A,
\end{equation*}
where $|u \rangle_{p, A} \in \mathcal{H}$ is a state that corresponds to a point $u^i \partial_i \in T_pM$. As discussed in section \ref{simplestcase}, the correct choice for $|u \rangle_{p, A} \in \mathcal{H}$ that reflects the vector-space structure of $T_pM$, is that of a coherent-state (\ref{cohstate1}). In order to make this state covariant with respect to the choice of the flat connection $A$, we define it through the property:
\begin{equation}
\hat{x}^T (g + i\omega) | u \rangle_{p, A} = u^T (g + i\omega) | u \rangle_{p, A},
\label{defup}
\end{equation}
that is $|u\rangle_{p,A}$ is the eigenstate of the annihilation operators $\hat{x}^i(g + i\omega)_{ij}$ defined according to the K\"ahler structure and coordinate system induced by the flat connection $A$. Clearly, this state is an element of $\mathcal{S}_d$, hence in the above defined wavefunction realization, our perturbation expansion (\ref{pertexp}) is completely well defined.

\medskip
\noindent
It is worth remarking here, that under the involution $J \rightarrow -J$, $g \rightarrow -g$, $\omega \rightarrow \omega$ or the involution $J \rightarrow J$, $g \rightarrow -g$, $\omega \rightarrow -\omega$ we would map a positive normed state to a ``negative normed state", which is therefore non-existent as an element of a (positive) Hilbert-space. We shall forget this remark until we encounter Lorentzian conic special K\"ahler manifolds in section \ref{conicsk}. Here, and until otherwise stated, we will restrict ourselves to Riemannian K\"ahler manifolds.

\medskip
\noindent
The fundamental property of $|u\rangle_{p,A}$, is that under a symplectomorphism with differential $\Sigma$ and corresponding unitary operator $S_\Sigma$, it transforms as follows:
\begin{equation}
|u \rangle_{p, \tilde{A}} \sim S_{\Sigma}|\Sigma^T u \rangle_{p, A},
\label{cohtang}
\end{equation}
where by $\sim$ we mean equal up to a phase and where $\tilde{A}$ is the gauge transformed connection (\ref{gaugetrans}). Equation (\ref{cohtang}) follows immediately from the fact that both the left- and right-hand side satisfy the defining equation (\ref{defup}) with $g + i\omega$ replaced by $\tilde{g} + i\tilde{\omega}$. This property translates to the following property for the wavefunction:
\begin{align}
Z_{\tilde{A}}(u, p) &= \, _{p, \tilde{A}} \!\,\langle u | p \rangle_{\tilde{A}}\nonumber\\
&= \, _{p, \tilde{A}} \!\,\langle u |S_{\Sigma} S_{\Sigma}^{-1}| p \rangle_{\tilde{A}}\nonumber\\
&= \, _{p, \tilde{A}} \!\,\langle u |S_{\Sigma} | p \rangle_A\nonumber\\
&= \, _{p, A} \!\,\langle \Sigma^T u | p \rangle_{A}\nonumber\\
&\sim Z_A(\Sigma^T u, p),
\label{Zastensor}
\end{align}
that is, $Z_A(u,p)$ should be viewed as a section of the line-bundle $\pi^*(\mathcal{L}\otimes \mathcal{L}' \,^{\vee}) \rightarrow TM$ where $\pi :TM \rightarrow M$ is the canonical projection of the tangent-bundle, $\mathcal{L}$ is the pullback, under the quantization map, of the tautological line-bundle on $\mathbb{P}^{\infty}$, and $\mathcal{L}'$ is a, at this point unspecified, unitary line bundle whose introduction is due to the fact that (\ref{cohtang}) is an ``equation up to a phase". Equivalently, property (\ref{Zastensor}) says that the object $Z_A(\cdot, p)$ is an element of $\Gamma(M, (\mathcal{L}\otimes \mathcal{L}' \,^{\vee})\otimes S^{\bullet}TM)$, which is what  we set out to achieve. By $S^{\bullet}$ we have denoted symmetric tensors.

\medskip
\noindent
Now we shall give an explicit expression for $|u \rangle_{p, \tilde{A}}$. To this aim we use the fact that for a K\"ahler manifold, if $\omega$ is in canonical form, the map transforming Darboux to Riemann normal coordinates erected at a point $p \in M$ is a symplectomorphism at $p$. Let's denote the differential from Riemann normal coordinates at $p$ by $\Lambda$ and corresponding gauge transformation by $S_{\Lambda}$, then:
\begin{equation}
|u \rangle_{p, A} \sim S_{\Lambda}|\Lambda^T u \rangle_{flat},\label{upatapoint}
\end{equation}
where by $| u \rangle_{flat}$, we denote the canonical, Poissonian coherent state (\ref{poisscohstate}).

\medskip
\noindent
For the following we will need to know how the Heisenberg algebra acts on $|u\rangle_{p, A}$ and also how $|u\rangle_{p, A}$ depends on $p$. Armed with the standard result in the flat case, under the assumption that the state is normalized to $1$, we obtain:
\begin{equation}
\hat{x} | u \rangle_{p, A} = \K{(}{)}{\frac{1}{2}(1 - iJ)^Tu + \frac{1}{2}(g^{-1} - i\omega^{-1})\vec{\nabla}_u + \frac{1}{4}(1 + iJ)^Tu} | u \rangle_{p, A}.
\label{defup2}
\end{equation}
It will be convenient in later sections to introduce the following notation:
\begin{align*}
v := \frac{1}{2}(1 - iJ)^T u\\
D_v := (1 - iJ) \vec{\nabla}_u.
\end{align*}
Thus equation (\ref{defup2}) takes the form:
\begin{equation*}
\hat{x} |u \rangle_{p, A} = \K{(}{)}{v + \frac{1}{2}g^{-1} D_v + \frac{1}{2} \overline{v}}|u \rangle_{p, A}.
\end{equation*}
We will also be needing the following simple identities. First of all:
\begin{equation*}
_{p, A} \! \,\langle 0 | u \rangle_{p, A} = \exp\K{(}{)}{-\frac{1}{4}||u||^2_{g(p)}}
\end{equation*}
and for any state $|\psi\rangle$, $\langle \psi | u \rangle_{p, A} $ is of the form:
\begin{equation*}
\langle \psi | u \rangle_{p, A} = \exp\K{(}{)}{-\frac{1}{4}||u||^2_{g(p)}} f_{\psi}((1 - iJ)^Tu),
\end{equation*}
where $f_{\psi}$ is an arbitrary, appropriately normalizable\footnote{We will discuss normalization conditions shortly.}, analytic function.
Moreover, if we assume that:
\begin{equation*}
_{p, A}\! \,\langle 0 | \psi \rangle \neq 0,
\end{equation*}
then we can write $f_{\psi}$ as the exponential of a power series in $u$.

\medskip
\noindent
Next we turn to the dependence of $|u \rangle_{p, A}$ on $p$. We shall thus analyse the action of the coordinate vectorfields at $p$ on $|u \rangle_{p, A}$:
\begin{align*}
\frac{\partial}{\partial x^k} | u \rangle_{p, A} &= \textrm{phase} \cdot \K{(}{)}{\frac{\partial}{\partial x^k} - i\beta_k}S_{\Lambda}| \Lambda^T u \rangle_{flat}\\
&= \K{(}{)}{-i\beta_k + (\partial_k S_{\Lambda})S_{\Lambda}^{-1} + (\partial_k \Lambda)_r^iu^r (\Lambda^{-1})_i^s \frac{\partial}{\partial u^s}} | u \rangle_{p, A}\\
&= \K{(}{)}{-i\beta_k + \frac{i}{2}(\Lambda(\partial_k \Lambda^{-1})\omega)_{ij}\hat{x}^i\hat{x}^j - u^r (\Lambda (\partial_k\Lambda^{-1}))_r^s \frac{\partial}{\partial u^s}} | u \rangle_{p, A}.
\end{align*}
In the above we have introduced a connection $i\beta = i\beta_k dx^k$ with $\beta_k \in \mathbb{R}$ on $\mathcal{L}'$. Now we use the fact that $\Lambda$ is related to the Levi-Civita connection through:
\begin{equation*}
\Gamma_k = -\Lambda \partial_k \Lambda^{-1}.
\end{equation*}
Of course the above is valid only at $p$, moreover, because of that, the connection $\beta_k$ should not be expected to be flat. Finally we obtain:
\begin{equation}
\K{(}{)}{\frac{\partial}{\partial x^k} + i\beta_k - u^r \Gamma_{kr}^s \frac{\partial}{\partial u^s} + \frac{i}{2}(\Gamma_k\omega)_{ij}\hat{x}^i\hat{x}^j} | u \rangle_p = 0
\label{defup3}
\end{equation}
More explicitly we have:
\begin{equation*}
\K{(}{)}{\frac{\partial}{\partial x^k} + i\beta_k - u^r \Gamma_{kr}^s \frac{\partial}{\partial u^s} + \frac{i}{2}(\Gamma_k\omega)_{ij}(v + \frac{1}{2}g^{-1} D_v + \frac{1}{2} \overline{v})^i(v + \frac{1}{2}g^{-1} D_v + \frac{1}{2} \overline{v})^j} | u \rangle_p = 0.
\end{equation*}
At this point it is important to notice that as we chose the map $\Lambda$ to Riemann normal coordinates we could have also chosen a vielbein\footnote{Of course this is valid only locally, however for parallelizable manifolds (e.g. Lie groups) this can hold globally.}. The difference reflects itself in the choice of connection $\beta$. More generally, the states $|u\rangle_p$ are parallel transported along $M$ (up to a phase) by the lift to the Hilbert bundle of any metric compatible and (quasi-)symplectic connection on $TM$. Choosing the vielbein instead of the map to Riemann normal coordinates corresponds to the Weitzenb\"ock connection \cite{weitzenboeck1}, which while not torsion free as the Levi-Civita connection, is flat. To see this we resort to the defining equation (\ref{defup}). Let $B$ denote a connection on $TM$. Choose a path $\gamma : [0,1] \rightarrow M$. The statement that $|u \rangle_p$ is parallel transported (up to a phase) by the lift of $B$ along $\gamma$ is:
\begin{equation*}
\K{|}{\rangle}{\K{[}{]}{\mathcal{P}\exp\K{(}{)}{\int_0^t \gamma^*B }}^Tu}_{\gamma(t)} \sim \mathcal{P}\exp\K{(}{)}{-\frac{i}{2} \int_0^t (\gamma^*B \omega)_{ij}\hat{x}^i \hat{x}^j} | u \rangle_{\gamma(0)}.
\end{equation*}
The above is equivalent to:
\begin{align*}
&(g(\gamma(t)) - i\omega)_{ij}\hat{x}^j \mathcal{P}\exp\K{(}{)}{\frac{i}{2} \int_0^t (\gamma^*B \omega)_{ij}\hat{x}^i \hat{x}^j} | u \rangle_{\gamma(0)}\\
&= (g(\gamma(t)) - i\omega)_{ij}\K{(}{)}{\K{[}{]}{\mathcal{P}\exp\K{(}{)}{\int_0^t \gamma^*B }}^Tu}^j \mathcal{P}\exp\K{(}{)}{\frac{i}{2} \int_0^t (\gamma^*B \omega)_{ij}\hat{x}^i \hat{x}^j}| u \rangle_{\gamma(0)},
\end{align*}
which infinitesimally reads:
\begin{align*}
0 &= X^k \K{(}{)}{\partial_k g_{ij} (\hat{x}^j - u^j) - (g - i\omega)_{ij} (B_k)_{l}^j (\hat{x}^l - u^l)}|u \rangle_0\\
&= X^k \K{(}{)}{(\partial_kg_{ij} - (B_k)_i^l g_{lj} - g_{il}(B_k)_j^l) + i ((B_k)_i^l \omega_{lj} + \omega_{il}(B_k)_j^l)}(\hat{x}^j - u^j) |u \rangle_0.
\end{align*}
We thus obtain that whatever the choice of $X$ and thus $\gamma$, the above is fulfilled provided $B$ is compatible with both metric and symplectic form.

\medskip
\noindent
To end this section we shall discuss the normalization condition on $Z_{A}(u,p)$. By this we mean the property that if $|p_0 \rangle_{A}$ is normalized to $1$ for a given point $p_0 \in M$ so will $|p \rangle_{A}$ for any other $p \in M$ due to the fact that the flat connection $A$ induces a unitary parallel transport. In order to express this property in terms of $Z_A(u,p)$, recall that in ordinary quantum mechanics, that is the quantum mechanics of $M = \mathbb{R}^{2d}$, the identity operator is expressed in terms of the coherent states as follows:
\begin{equation}
\mathrm{Id} = \frac{1}{(2\pi)^d} \int_{\mathbb{R}^{2n}} \; |u \rangle \langle u | \; du^1 \wedge \dots \wedge du^{2d}.\label{norm1}
\end{equation}
Thus, resorting to (\ref{upatapoint}), in our case we obtain:
\begin{equation}
\mathrm{Id} = \frac{1}{(2\pi)^d} \int_{T_pM} \sqrt{\det(g_A(p))} \; |u \rangle_{p, A} \; _{p,A}\!\,\langle u | \; du^1 \wedge \dots \wedge du^{2d},\label{norm2}
\end{equation}
and therefore:
\begin{equation}
1 = \frac{1}{(2\pi)^d} \int_{T_pM} \sqrt{\det(g_A(p))} \; \K{|}{|}{Z_A(u,p)}^2 \; du^1 \wedge \dots \wedge du^{2d}.
\label{norm3}
\end{equation}
In the above we have denoted the metric by $g_A$ to emphasize that it is expressed in the coordinate system corresponding to $A$.
\subsection{Master equation}\label{MEsection}
At this point we have all the ingredients to formulate the master equation. By this we mean the statement that $|p\rangle_A$ is parallel transported by the flat connection $A$, expressed as a differential equation for the wavefunction $Z_A (u, p)$. Using (\ref{defup3}) we obtain:
\begin{align*}
0 &=  \,  _{p, A} \! \,\langle u | \frac{\partial}{\partial x^k} + A_k | p \rangle_A\\
&= \frac{\partial}{\partial x^k}  Z_A(u,p) - \frac{\partial}{\partial x^k} \K{(}{)}{_{p, A} \! \,\langle u |}  | p \rangle_A + \, _{p, A} \! \,\langle u | A_k | p \rangle_A\\
&= \K{(}{)}{\frac{\partial}{\partial x^k} - i\beta_k  - u^r \Gamma_{kr}^s \frac{\partial}{\partial u^s}}Z_A(u,p) + \, _{p, A} \! \,\langle u | - \frac{i}{2}(\Gamma_k\omega)_{ij}\hat{x}^i\hat{x}^j + A_k | p \rangle_A\\
&= \K{(}{)}{\frac{\partial}{\partial x^k} - i\beta_k - u^r \Gamma_{kr}^s \frac{\partial}{\partial u^s}}Z_A(u,p) + i\, _{p, A} \! \,\langle u |\alpha_k + \omega_{ik}\hat{x}^i - \frac{1}{2}((\Gamma_k - G_k) \omega)_{ij}\hat{x}^i \hat{x}^j | p \rangle_A.
\end{align*}
Resorting to (\ref{defup2}) we thus obtain:
\begin{align}
&\K{(}{.}{\frac{\partial}{\partial x^k}  - u^r \Gamma_{kr}^s \frac{\partial}{\partial u^s} + i (\alpha_k - \beta_k) - i \omega_{ki}(\overline{v} + \frac{1}{2}g^{-1} \overline{D}_v + \frac{1}{2} v)^i}\nonumber\\
&\K{.}{)}{+ \frac{i}{2}C_{kij}(\overline{v} + \frac{1}{2}g^{-1} \overline{D}_v + \frac{1}{2} v)^i(\overline{v} + \frac{1}{2}g^{-1} \overline{D}_v + \frac{1}{2} v)^j} Z_A(u, p) = 0,\label{masterequation}
\end{align}
where the tensor $C$ is given by:
\begin{equation*}
C_{kij} = ((\Gamma_k - G_k)\omega)_{ij}.
\end{equation*}
Equation (\ref{masterequation}) should be viewed as a version of the holomorphic anomaly equation of \cite{bcov9309140}, which is the master equation for the generating function of topological closed-string amplitudes. However at this point this statement is not completely transparent. Indeed in our case the phase space is an affine special K\"ahler manifold, while what should play the role of a classical phase space in topological string theory is the vector-multiplet moduli space of $\mathcal{N}=(2,2)$ conformal field theories. This has the structure of a projective special K\"ahler manifold. We shall tackle this geometry in section \ref{projsk}. The crucial point is that projective special K\"ahler manifolds can be recovered as quotients of affine conic (however Lorentzian) special K\"ahler manifolds.

\medskip
\noindent
A further remark, that we will clarify in later sections, is that so far Planck's constant $\hbar$ has not manifestly appeared in our quantization scheme. At this point, its introduction would be merely as an arbitrary rescaling of the symplectic form. We will instead see in section \ref{projsk} how the notion of Planck's constant arises naturally in the passage from affine to projective geometry. 

\medskip
\noindent
Before delving into these matters we will analyze the solution to the master equation. We shall denote by $|p\rangle^s$ the coherent state of $p \in M$ and $|u\rangle_p^s$ the coherent state of $u \in T_pM$ in special Darboux coordinates gauge. Then, the master equation reduces to:
\begin{align*}
&\K{(}{.}{\frac{\partial}{\partial x^k} - u^r \Gamma_{kr}^s \frac{\partial}{\partial u^s} + i(\alpha_k - \beta_k) - i\omega_{ki}(\overline{v}  + \frac{1}{2} v)^i + \frac{1}{2}(\overline{D}_v)_k}\\
&\K{.}{)}{+ \frac{i}{2}(\Gamma_k\omega)_{ij}\overline{v}^i\overline{v}^j + \frac{i}{8}(\Gamma_k\omega)_{ij}(g^{-1} \overline{D}_v + v)^i(g^{-1} \overline{D}_v + v)^j}Z_s(u,p) = 0,
\end{align*}
where we have used $\omega_{ki}(g^{-1}\overline{D}_v)^i = i (\overline{D})_k$ and the fact that $\Gamma_k \omega$ splits into holomorphic and anti-holomorphic components. This is immediately verified using the explicit formula:
\begin{equation*}
(\Gamma_k \omega)_{ij} = \frac{1}{2} \partial_i \partial_k \partial_r K J_j^r.
\end{equation*}
Since $C$ is a tensor, it will split in holomorphic and anti-holomorphic components in any coordinate system. Thus the equation above is valid in general with $(\Gamma_k\omega)_{ij}$ replaced by $C_{kij}$. It will be convenient in the following to split the master equation into its holomorphic and anti-holomorphic parts. In the following we shall compute in special Darboux coordinates. The anti-holomorphic part is then given by:
\begin{align*}
&(1 + iJ)_l^k \K{(}{.}{\frac{\partial}{\partial x^k} - \frac{1}{2}\overline{v}^r \Gamma_{kr}^s (D_v)_s +i(\alpha_k - \beta_k) - \frac{i}{2}\omega_{ki} v^i}\\
&\quad \quad \quad \quad \quad \K{.}{)}{ + \frac{1}{2}(\overline{D}_v)_k + \frac{i}{2}C_{kij}\overline{v}^i\overline{v}^j}Z_s(u,p) = 0,
\end{align*}
while the holomorphic part reads:
\begin{align*}
&(1 - iJ)_l^k \K{(}{.}{\frac{\partial}{\partial x^k} - \frac{1}{2}v^r \Gamma_{kr}^s (\overline{D}_v)_s +i(\alpha_k - \beta_k) - i\omega_{ki}\overline{v}^i}\\
&\quad \quad \quad \quad \quad \K{.}{)}{ + \frac{i}{8}C_{kij}(g^{-1} \overline{D}_v + v)^i(g^{-1} \overline{D}_v + v)^j}Z_s(u,p) = 0.
\end{align*}
We shall now assume:
\begin{equation}
_{p, A} \! \,\langle 0 | p \rangle_{A} \neq 0.
\label{nonzerosection}
\end{equation}
This is no loss of generality as long as one restricts attention to a small enough neighborhood of $p$. Then as discussed in section \ref{CTB}, we are allowed to write the following ansatz for $Z_s(u,p)$:
\begin{equation*}
Z_s(u, p) = \exp\K{(}{)}{\sum_{n \geq 0}\sum_{i_1, \dots, i_n} \frac{1}{n!} C^n_{i_1, \dots, i_n}\overline{v}^{i_1}\cdots \overline{v}^{i_n} - \frac{1}{4}||u||^2_{g(p)}},
\end{equation*}
where the $C^n$'s are symmetric tensors. For the following we shall need a few identities:
\begin{align*}
(\overline{D}_v)_k v^l &= 0\\
(\overline{D}_v)_k \overline{v}^l &= (1 + iJ)_k^l\\
||u||_g^2 &= 2\overline{v}^i v^j g_{ij}.
\end{align*}
First we analyze the equation arising from the anti-holomorphic part of the flat connection:

\medskip
\noindent
\begin{align}
0 &= \sum_{n \geq 0}\sum_{i_1, \dots, i_n} \frac{1}{n!} \overline{v}^k \partial_k C^n_{i_1, \dots, i_n}\overline{v}^{i_1}\cdots \overline{v}^{i_n}\nonumber\\
&+ \frac{1}{n!} \overline{v}^k \sum_{r = 1}^n C^n_{i_1, \dots, i_n}\overline{v}^{i_1} \cdots \frac{i}{2}\partial_k J_s^{i_r} u^s \cdots \overline{v}^{i_n} \label{Jsub}\\
&+ \frac{1}{(n-1)!} C^n_{i_1, \dots, i_n}\overline{v}^{i_1} \cdots \overline{v}^{i_n}\nonumber\\
& + i\overline{v}^k(\alpha_k - \beta_k) +  \frac{i}{2}C_{kij}\overline{v}^k\overline{v}^i\overline{v}^j.\nonumber
\end{align}
Now we use the fact that in a K\"ahler manifold, the complex structure is parallel transported by the Levi-Civita connection. In components this reads:
\begin{equation*}
\partial_k J_s^r + J_s^t \Gamma_{kt}^r - J_t^r \Gamma_{ks}^t = 0.
\end{equation*}
We substitute for $\partial_k J_s^{i_r}$ in (\ref{Jsub}). The part of the expression involving the Levi-Civita connection thus becomes:
\begin{align*}
&\frac{i}{2n!}\overline{v}^k \sum_{r = 1}^n C^n_{i_1, \dots, i_n} \overline{v}^{i_1} \dots \K{[}{]}{-J_s^t\Gamma_{kt}^{i_r} + J_t^{i_r} \Gamma_{ks}^t}u^s \cdots \overline{v}^{i_n}\\
&= - \frac{1}{n!} \overline{v}^k \Gamma_{ks}^{i_r} \overline{v}^s\sum_{r = 1}^n C^n_{i_1, \dots, i_n} \overline{v}^{i_1} \dots \widehat{\overline{v}^{i_r}} \cdots \overline{v}^{i_n}\\
&= 0.
\end{align*}
We thus obtain the following recursive formula:
\begin{equation*}
C^{n+1}_{i_1, \dots, i_{n+1}} \overline{v}^{i_1} \cdots \overline{v}^{i_{n+1}} = -\overline{v}^k\partial_k C^{n}_{i_1, \dots, i_n} \overline{v}^{i_1} \cdots \overline{v}^{i_n} 
\end{equation*}
for all $n \geq 3$, while the lower terms yield:
\begin{align*}
C^1_{i_1} \overline{v}^{i_1} &= -\overline{v}^k \partial_k C^0 - i\overline{v}^k (\alpha_k - \beta_k)\\
C^2_{i_1, i_2} \overline{v}^{i_1} \overline{v}^{i_2} &= -\overline{v}^k \partial_k C^1_{i_1} \overline{v}^{i_1}\\
C^3_{i_1, i_2, i_3}\overline{v}^{i_1}\overline{v}^{i_2}\overline{v}^{i_3} &= -\overline{v}^k\partial_k C^{2}_{i_1, i_2} \overline{v}^{i_1} \overline{v}^{i_2} - iC_{ijk}\overline{v}^i\overline{v}^j\overline{v}^k.
\end{align*}
At this point it is convenient to introduce complex coordinates $(z^1, \dots, z^d)$ with $d = \mathrm{dim}(M)/2$ that we shall label with greek letters. We shall further denote by $\nabla^{(0,1)}$ the anti-holomorphic part of the Levi-Civita covariant derivative. We now introduce the covariant tensors $\mathcal{C}^n$ defined by:
\begin{align*}
\mathcal{C}^n &= \frac{(-1)^n}{2^n}C_{i_1, \dots, i_n}(1 + iJ)_{j_1}^{i_1} \cdots (1 + iJ)_{j_n}^{i_n} dx^{j_1}\cdots dx^{j_n}\\
&= \mathcal{C}^n_{\overline{\mu}_1, \dots, \overline{\mu}_n} dz^{\overline{\mu}_1}\cdots dz^{\overline{\mu}_n},
\end{align*}
where by ``$\cdot$" we denote the symmetrized tensor product. The above equations then take the form:
\begin{align}
\mathcal{C}^1 &= \overline{\partial} \mathcal{C}^0 + i(\alpha_{\overline{\mu}} - \beta_{\overline{\mu}})dz^{\overline{\mu}}\label{rec1}\\
\mathcal{C}^3 &= \nabla^{(0, 1)} \mathcal{C}^2 + iC_{\overline{\mu}_1, \overline{\mu}_2, \overline{\mu}_3}dz^{\overline{\mu}_1}dz^{\overline{\mu}_2}dz^{\overline{\mu}_3}\\
\mathcal{C}^{n+1} &= \nabla^{(0, 1)} \mathcal{C}^n \quad \forall n \geq 1, \; n \neq 2.\label{rec3}
\end{align}
Notice, in particular, that the solution to the master equation is completely determined by a single object, $\mathcal{C}^0$. Clearly the latter is given by:
\begin{equation*}
_p \!\, \langle 0 | p \rangle = \exp(\mathcal{C}^0(p)) =: c(p).
\end{equation*}
Notice, that thanks to property (\ref{Zastensor}), this quantity is independent of $A$. Indeed the above is a section of $\mathcal{L} \otimes \mathcal{L}'\,^{\vee} \rightarrow M$ which (by assumption (\ref{nonzerosection})) is non-vanishing over the open set under consideration. This section is, in turn, determined by the holomorphic part of the master equation.
For its analysis, it is convenient to replace the wavefunction by the inhomogeneous tensor:
\begin{equation*}
\mathcal{C} = \exp\K{(}{)}{\sum_{n \geq 0} \frac{(-1)^n}{n!}\mathcal{C}^n}.
\end{equation*}
In terms of this, the anti-holomorhic part of the master equation reads:
\begin{equation}
\K{(}{)}{\nabla^{(0,1)} + i(\alpha_{\overline{\mu}} - \beta_{\overline{\mu}})dz^{\overline{\mu}} + d\overline{z}^{\overline{\mu}}\iota_{\partial_{\overline{\mu}}} + \frac{i}{2}C_{\overline{\mu}\overline{\nu}\overline{\rho}}d\overline{z}^{\overline{\mu}}d\overline{z}^{\overline{\nu}}d\overline{z}^{\overline{\rho}}}\mathcal{C} = 0.\label{masterantiholo1}
\end{equation}
The holomorphic part instead becomes:
\begin{equation}
\K{(}{)}{\partial + \frac{i}{2}dz^{\mu} C_{\mu \nu \rho} g^{\nu \overline{\nu}} g^{\rho \overline{\rho}} \iota_{\partial_{\overline{\nu}}} \iota_{\partial_{\overline{\rho}}} + i(\alpha_{\mu} - \beta_{\mu})dz^{\mu} - \frac{i}{2}\omega}\mathcal{C} = 0.\label{masterholo1}
\end{equation}
The above can be seen as yielding an infinite number of differential equations for the section $c$. 

\medskip
\noindent
Now we shall analyse the integrability of the master equation. As we will see, and as is to be expected, this precisely specifies the line-bundle $\mathcal{L}'$. The result of this computation are the following three equations, namely the $(0,2)$, $(1,1)$ and $(2,0)$ components of the underlying Maurer-Cartan equation respectively:
\begin{align*}
\partial_{\overline{\mu}} (\alpha_{\overline{\nu}} - \beta_{\overline{\nu}}) - \partial_{\overline{\nu}} (\alpha_{\overline{\mu}} - \beta_{\overline{\mu}}) &= 0\\
i\partial_{\mu} (\alpha_{\overline{\nu}} - \beta_{\overline{\nu}}) - i\partial_{\overline{\nu}}(\alpha_{\mu} - \beta_{\mu}) &=\frac{1}{2}\K{(}{)}{C_{\mu \rho \sigma}g^{\rho \overline{\rho}} g^{\sigma \overline{\sigma}} C_{\overline{\nu}\overline{\rho}\overline{\sigma}} + 2g_{\mu \overline{\nu}}}\\
\partial_{\mu}(\alpha_{\nu} - \beta_{\nu}) - \partial_{\nu}(\alpha_{\mu} - \beta_{\mu}) &= 0.
\end{align*}
In particular, choosing $\alpha$ in canonical form (\ref{canonalpha}), we have:

\medskip
\noindent
\begin{align*}
\partial_{\overline{\mu}} \beta_{\overline{\nu}} - \partial_{\overline{\nu}} \beta_{\overline{\mu}} &= 0\\
i\partial_{\mu} \beta_{\overline{\nu}} - i\partial_{\overline{\nu}}\beta_{\mu} &=-\frac{1}{2}C_{\mu \rho \sigma}g^{\rho \overline{\rho}} g^{\sigma \overline{\sigma}} C_{\overline{\nu}\overline{\rho}\overline{\sigma}} = -\frac{1}{2}R_{\mu \overline{\nu}} = \frac{i}{2}\rho_{\mu \overline{\nu}}\\
\partial_{\mu}\beta_{\nu} - \partial_{\nu}\beta_{\mu} &= 0,
\end{align*}
where $R_{\mu \overline{\nu}}$ and $\rho_{\mu \overline{\nu}}$ denote the components of the Ricci tensor and Ricci form respectively. See appendix \ref{appendix} for the identity used in the second equation. Thus, since smooth line-bundles are completely specified by their first Chern-class we obtain that $\mathcal{L}'$ is isomorphic, as a smooth bundle, to the square-root of the canonical bundle. Moreover $\beta$ is in canonical form:
\begin{equation*}
\beta = i(\partial - \overline{\partial})\chi
\end{equation*}
and up to holomorphic gauge transformations, we can choose:
\begin{equation*}
\chi = \frac{1}{4}\log \sqrt{\mathrm{det}g}.
\end{equation*}
We can trivially twist the line-bundle $\mathcal{L}\otimes \mathcal{L}'^{\vee}$ to an anti-holomorphic line-bundle by multiplying $\mathcal{C}$ by an appropriate factor. We thus define:
\begin{equation*}
\mathcal{S} := \K{(}{)}{\mathrm{det}\,g}^{\frac{1}{8}}e^{\frac{K}{2}}\mathcal{C}.
\end{equation*}
Finally, in terms of $\mathcal{S}$, the master equation acquires the form of a ``(anti-)\\
holomorphic-anomaly equation":
\begin{align}
\K{(}{)}{\nabla^{(0,1)} - \overline{\partial}K - \frac{1}{2}\overline{\partial}\log \sqrt{\det g}+ d\overline{z}^{\overline{\mu}}\iota_{\partial_{\overline{\mu}}} + \frac{i}{2}C_{\overline{\mu}\overline{\nu}\overline{\rho}}d\overline{z}^{\overline{\mu}}d\overline{z}^{\overline{\nu}}d\overline{z}^{\overline{\rho}}}\mathcal{S} &= 0
\label{holoaneq1}\\
\K{(}{)}{\partial + \frac{i}{2}dz^{\mu} C_{\mu \nu \rho} g^{\nu \overline{\nu}} g^{\rho \overline{\rho}} \iota_{\partial_{\overline{\nu}}} \iota_{\partial_{\overline{\rho}}} - \frac{i}{2}\omega}\mathcal{S} &= 0.
\label{holoaneq2}
\end{align}
Now, we shall consider the section of the holomorphic line-bundle $(\mathcal{L}\otimes \mathcal{L}' \,^{\vee})_{ahol}$:
\begin{equation*}
s(p) := \exp(\mathcal{S}^0).
\end{equation*}
The first order component of the holomorphic part of the master equation reads:
\begin{equation}
\partial_{\mu} s = -\frac{i}{2}C_{\mu \nu \rho}g^{\nu \overline{\nu}}g^{\rho \overline{\rho}}\K{(}{)}{D^{(0,1)} D^{(0,1)} s}_{\overline{\nu}\overline{\rho}},\label{holoanomaly}
\end{equation}
where $D^{(0,1)}$ denotes the anti-holomorphic part of the Levi-Civita connection twisted by the Chern connection on $(\mathcal{L}\otimes \mathcal{L}' \,^{\vee})_{ahol}$, which is naturally a hermitian line-bundle with hermititan form:
\begin{equation*}
h = (\det g)^{-\frac{1}{4}}e^{-K}.
\end{equation*}
In fact, the integrability of the master equation ensures that a solution $s$ to (\ref{holoanomaly}) lifts, through the recursion relations (\ref{rec1}--\ref{rec3}), to a solution of the full master equation. Of course, a priori, there are, if any, more than one solution to the above equation. indeed in the simplest case, i.e. $M = \mathbb{R}^{2d}$, there are infinitely many solutions. This follows immediately from the fact that, in that particular case, the tensor $C$ vanishes and thus any anti-holomorphic section $s$ solves the problem. Recall that this arbitrariness corresponds to the freedom of choosing the image of a particular marked point on $M$ via the quantization map. The canonical quantization of $\mathbb{R}^{2d}$ has as solution $s = 1$. In the next subsection we will construct the general solution for any affine special K\"ahler manifold.
\subsection{Constructing the solution I: the role of special coordinates}\label{sol1}
In this section we will give an explicit expression for the Green's function of the holomorphic anomaly equation, thereby providing its general solution. We start by choosing an arbitrary point $p_0 \in M$ and declare that:
\begin{equation*}
p_0 \stackrel{\phi}\mapsto |p_0 \rangle_A := |y\rangle_{p_0, A}.
\end{equation*}
Now we ask where a point $p$, in its Darboux neighborhood, is mapped to. For this we shall consider the action of the annihilation operators on $|p \rangle$
\begin{align*}
\hat{x}^T(g + i\omega)_{p_0, A}|p \rangle_A &= \hat{x}^T(g + i\omega)_{p_0, A}U(p, p_0)|y \rangle_{p_0, A}\\
&= U_A(p, p_0)\K{(}{)}{U_A(p_0, p) \hat{x}^T U_A(p, p_0)}(g + i\omega)_{p_0, A}|y \rangle_{p_0, A}.\\
\end{align*}
The $p$ dependent operator $\hat{x}^T$ is found by computing its infinitesimal variation:
\begin{align*}
\frac{\partial}{\partial x^k}\K{(}{|}{U_A(p, p')\hat{x}^lU_A(p', p)}_{p' = p} &= [A_k, \hat{x}^l]\\
&= i[\omega_{ik}\hat{x}^i - D_{ijk}\hat{x}^i \hat{x}^j, \hat{x}^l]\\
&= -\omega_{ik}\omega^{il}  -2D_{ijk}\omega^{jl}\hat{x}^i\\
&= \delta_k^l - G_{k i}^l \hat{x}^i.
\end{align*}
Bringing the last term to the left hand side we thus obtain:
\begin{equation*}
(\partial_k + G_k)\K{(}{|}{U_A(p, p')\hat{x}U_A(p', p)}_{p' = p} = \delta_k^l,
\end{equation*}
therefore:
\begin{equation*}
U_s(p_0, p)\hat{x}U_s(p, p_0) = \hat{x} + x_{p_0}^p,
\end{equation*}
where $x_{p_0}^p$ is the coordinate vector of $p$ in special Darboux coordinates around $p_0$. Therefore:
\begin{equation*}
\hat{x}^T(g + i\omega)_{p_0, s}|p \rangle_s = (x_{p_0}^p + y)^T (g + i\omega)_{p_0}|p \rangle_{s},
\end{equation*}
hence:
\begin{equation*}
|p \rangle_s \sim |x_{p_0}^p + y\rangle_{p_0, A_s}.
\end{equation*}
From the point of view of topological string theory, what the above equation means is that the topological conformal field theory corresponding to the point $p$ is related to the one at $p_0$ by a deformation whose modulus corresponds to the special coordinate vector of $p$ relative to $p_0$.

\medskip
\noindent
Now we shall fix the phase ambiguity. We will need the following identity:
\begin{equation*}
\K{(}{)}{(\overline{D}_v)_k + \frac{1}{2}(g - i\omega)_{kj}u^j}|u \rangle_p = 0
\end{equation*}
and for simplicity, until otherwise stated we shall denote $x_{p_0}^p$ simply by $x$. Then
\begin{align*}
0 &= \K{(}{)}{\frac{\partial}{\partial x^k} + i(\alpha_k - \gamma_k) - i\omega_{ki}\hat{x}^i}|x + y\rangle_{p_0, A_s}\\
&= \K{(}{)}{\frac{\partial}{\partial x^k} + i(\alpha_k - \gamma_k) - i\omega_{ki}(v + \frac{1}{2}g_{p_0}^{-1}D_v + \frac{1}{2}\overline{v})^i}|x + y\rangle_{p_0, A_s}\\
&= i(\alpha_k - \gamma_k - \frac{1}{2}\omega_{kj}(x + y)^j)|x + y\rangle_{p_0, A_s},
\end{align*}
where $i\gamma_k = \partial_k \theta$ and $\theta$ is the phase discrepancy. Also, in the present case, $v = (1/2)(1 - iJ)x$. Thus, since we have chosen $\alpha$ in canonical form:
\begin{align*}
\gamma_k &= \alpha_k - \frac{1}{2}\omega_{kj}(x + y)^j\\
&= -\frac{1}{2}J_k^l \partial_lK - \frac{1}{2}\omega_{kj}(x + y)^j.
\end{align*}
So we have:
\begin{equation*}
|p \rangle_s = \exp\K{(}{)}{\frac{i}{2}\int_{p_0}^p (J_k^l \partial_l K + \omega_{kj}(x + y)^j) dx^k}|x + y\rangle_{p_0, A_s}.
\end{equation*}
Now we are left to compute the kernel:
\begin{equation}
K(u, p; x + y, p_0) := \; _{p, A_s} \! \, \langle u | p\rangle_s^y.\label{propagator1}
\end{equation}
Given the above, the general solution to the master equation over a special Darboux patch is given by:

\medskip
\noindent
\begin{align}
Z(u, p)^f = \int_{T_{p_0}M} &\sqrt{\det g_s(p_0)} \, dy^1 \wedge \cdots \wedge dy^{2d}\, K(u, p, x_s + y, p_0)\cdot\\
&\exp\K{(}{)}{-\frac{1}{4}||y||_{g_s(p_0)}^2}f((1 + iJ_0)y),\label{generalsolution1}
\end{align}
where $f$ is any analytic gaussian-integrable function. The final step is thus to obtain an expression for the states $|u\rangle_{p,s}$ that is valid over an entire Darboux patch, rather than just at a point as we previously defined them in (\ref{upatapoint}).
\subsection{Constructing the solution II: revisiting the coherent tangent bundle}\label{sol2}
In this section we will give a patchwise description of the coherent states $|u \rangle_{p,s}$. The simplest way to find $|u\rangle_p$ is through the defining differential equations (\ref{defup}, \ref{defup2}, \ref{defup3}). It is convenient to express $|u \rangle_{p,s}$ as the wavefunction $\psi_u(q, J) = \langle q | u \rangle_{p,s}$, where $J$ is the complex structure matrix at the point $p$ and $|q\rangle$ is the eigenstate of $\hat{q}$ with vector of eigenvalues $q$. Here the vector of operators $\hat{q}$ is the upper half of the vector $\hat{x}$ in special Darboux coordinates, that is the position coordinates, rather than the momentum coordinates, which we will not label to avoid confusion with the label $p$ that stands for the point on the manifold on which these wavefunctions are erected. Accordingly we will need to split metric and symplectic form in $d \times d$ blocks:
\begin{align*}
g =:  \K{(}{)}{\begin{array}{cc} R_1 & R_2 \\ R_2^T & R_4 \end{array}}, \s \omega =  \K{(}{)}{\begin{array}{cc} 0 & -1 \\ 1 & 0 \end{array}}.
\end{align*}
Then (\ref{defup}) becomes:
\begin{equation*}
\K{(}{)}{\begin{array}{cc} R_1 & R_2 + i \\ R_2^T - i & R_4 \end{array}}\K{(}{)}{\begin{array}{cc} q - u_q \\ -i \nabla_q - u_p \end{array}} \psi_u(q, J) = 0,
\end{equation*}
which is equivalent to:
\begin{equation}
\nabla_q \psi_u = (i \tau (q - u_q) + iu_p)\psi_u,\label{cohstatewf2}
\end{equation}
where
\begin{equation*}
\tau = R_4^{-1}(i - R_2^T) = -(R_2 + i)^{-1}R_1
\end{equation*}
is a symmetric matrix called the complex modulus and encodes one-to-one the complex structure $J(p)$. In particular $\tau$ is an element of the Siegel upper-half space \cite{mumford-theta3}, which we shall denote as $\mathbb{H}_d$. A consequence of which is that:
\begin{equation*}
\det \mathrm{Im} \tau > 0.
\end{equation*}
The solution to (\ref{cohstatewf2}) is then:
\begin{equation*}
\psi_u(q, \tau) = \mathcal{N}(\tau, u) \exp \K{(}{)}{\frac{i}{2}\langle (q - u_q), \tau (q - u_q)\rangle + i u_p q},
\end{equation*}
where $\mathcal{N}(\tau, u)$ is yet to be determined. The latter is however constrained by three conditions. The first is the normalization condition on $|u \rangle_{p,s}$ while the second and third are the defining equations (\ref{defup2}, \ref{defup3}). The first implies:
\begin{equation*}
\mathcal{N}(\tau, u) = |\mathcal{N}(\tau, u)| e^{i\theta(\tau, u)}
\end{equation*}
where:
\begin{align*}
|\mathcal{N}| &= \K{(}{)}{\int_{\mathbb{R}^d} d^d q \K{|}{|}{\exp \K{(}{)}{\frac{i}{2}\langle (q - u_q), \tau (q - u_q)\rangle + i u_p q}}^2}^{-\frac{1}{2}}\\
&= \pi^{-d/4} \K{(}{)}{\mathrm{det} \, \mathrm{Im} \tau}^{1/4}.
\end{align*}
Equation (\ref{defup2}) then reduces to:
\begin{equation*}
(\nabla_{u_q} + \overline{\tau}\nabla_{u_p}) \theta  = -\frac{1}{2}(u_p + \overline{\tau}u_q).
\end{equation*}
Solving for the real and imaginary parts separately we obtain:
\begin{equation*}
\theta = -\frac{1}{2}\langle u_q, u_p\rangle + \gamma(\tau).
\end{equation*}
Thus we arrive at the following solution:
\begin{equation*}
\psi_u(q, \tau) = \pi^{-d/4} \K{(}{)}{\mathrm{det} \, \mathrm{Im} \tau}^{1/4}\exp\K{(}{)}{i\gamma(\tau) -\frac{i}{2}\langle u_q, u_p \rangle + \frac{i}{2}\langle (q - u_q), \tau (q - u_q) \rangle + i u_p \, q},
\end{equation*}
where the only undetermined quantity is the phase $\gamma(\tau)$. The phase is fixed (always up to an irrelevant constant), by equation (\ref{defup3}) and the choice of the connection $\beta$ on $\mathcal{L}'$. In particular equation (\ref{defup3}) yields:
\begin{equation*}
\partial_k \gamma = -\beta_k - \frac{1}{2}\mathrm{tr}\K{(}{)}{[(\Gamma_k \omega)_{pp}] \mathrm{Im} \tau}.
\end{equation*}
Here we remark the similarity of $\psi_u(q, \tau)$ with the wavefunctions discussed in \cite{dvv}. At this point we have all the ingredients to compute the kernel (\ref{propagator1}) of the master equation. We start with the computation of the overlap between a coherent state at $p_0$, where the complex modulus is $\tau_1$, and one at $p$ with complex modulus $\tau_2$:
\begin{align*}
&_{\tau_2} \! \, \langle u_2 | u_1 \rangle_{\tau_1} = \int d^d q \overline{\psi}(q, u_2, \tau_2) \psi(q, u_1, \tau_1) =\\
& (2i)^{d/2}\frac{(\det \mathrm{Im} \tau_1)^{1/4} (\det \mathrm{Im} \tau_2)^{1/4}}{(\det (\tau_1 - \overline{\tau_2}))^{1/2}} \cdot\\
& \exp \K{(}{)}{ - \frac{i}{2}\langle u_{q,1}, z_1 \rangle + \frac{i}{2} \langle u_{q,2}, \overline{z}_2\rangle - \frac{i}{2} \langle (z_1 - \overline{z}_2), (\tau_1 - \overline{\tau}_2)^{-1} (z_1 - \overline{z}_2)\rangle}\\
& \cdot \exp\K{(}{)}{i \int_{p_0}^{p} \K{(}{)}{\beta + \frac{1}{2} \mathrm{tr} ((\Gamma_k \omega)_{pp} \mathrm{Im} \tau) dx^k}},
\end{align*}
where we have introduced the complex coordinates $z = u_p - \tau u_q$. We shall now denote by $z_1$ the coordinates corresponding to $u_1 = x + y$ and by $z_2$ the ones corresponding to $u$, then the kernel is given by:
\begin{align*}
&K(u, p; x + y, p_0) = (2i)^{d/2}\frac{(\det \mathrm{Im} \tau_1)^{1/4} (\det \mathrm{Im} \tau_2)^{1/4}}{(\det (\tau_1 - \overline{\tau_2}))^{1/2}} \cdot \\
& \exp\K{(}{)}{-i \int_{p_0}^{p} \K{(}{)}{(\alpha - \beta) - \frac{1}{2}\omega_{kj}(x + y)^j dx^k - \frac{1}{2} \mathrm{tr} ((\Gamma_k \omega)_{pp} \mathrm{Im} \tau) dx^k}}\\
&\exp \K{(}{)}{ - \frac{1}{4}||u||_{g(p)}^2 - \frac{i}{2}\langle u_{q,1}, z_1 \rangle + \frac{1}{4} \langle \overline{z}_2, R_4(p) \overline{z}_2\rangle - \frac{i}{2} \langle (z_1 - \overline{z}_2), (\tau_1 - \overline{\tau}_2)^{-1} (z_1 - \overline{z}_2)\rangle}.
\end{align*}
We thus conclude the study of affine Riemannian special K\"ahler manifolds having provided the general solution (\ref{generalsolution1}) to the master equation (\ref{masterequation}).
\section{Conic special K\"ahler manifolds}\label{conicsk}
First of all we shall recover the structure of a projective special K\"ahler manifold in a way best suited for our quantization technique. Before that we shall recall the standard definitions (see e.g. \cite{specgeoreview} for a comprehensive review). 
\begin{definition}
(\textbf{Projective special K\"ahler manifold - 1}) A projective special K\"ahler manifold is a holomorphic quotient of an affine conic special K\"ahler manifold by its defining holomorphic $\mathbb{C}^*$ action.
\end{definition}
\begin{definition}\label{defconic1}
(\textbf{Conic special K\"ahler manifold - 1}) An affine conic special K\"ahler manifold is an affine special K\"ahler manifold equipped with a free holomorphic $\mathbb{C}^*$ action whose generating holomorphic vectorfield $H$ is a homothetic Killing vectorfield for the flat symplectic connection:
\begin{equation}
(d + G)H = \pi^{(1,0)} := \frac{1}{2}(1 - iJ)_k^l \, dx^k \otimes \partial_l.\label{homotheticvf}
\end{equation}
\end{definition}
Recall, that by definition of affine special K\"ahler: $d_{d + G} \pi^{(1,0)} = 0$. This ensures the existence of a vectorfield that satisfies the equation above. The restriction here, is that this vectorfield is required to be holomorphic.

\medskip
\noindent
We shall now analyse equation (\ref{homotheticvf}) in special Darboux coordinates. We shall introduce the vectorfield $X$ through:
\begin{equation*}
H = \pi^{(1, 0)} (-iX) = -\frac{i}{2}(1 - iJ) X.
\end{equation*}
Without loss of generality, we can choose $X$ real. Then equation (\ref{homotheticvf}) becomes:
\begin{align}
\partial_i (- JX)^j &= \delta_i^j\label{homotheticvf1}\\
\partial_i X^j = J_i^j.
\end{align}
Thus, up to an irrelevant constant vectorfield, from the second equation we obtain:
\begin{equation*}
X^j = \omega^{rj}\partial_rK,
\end{equation*}
which can be recast in the form
\begin{equation*}
dK = \iota_X \omega,
\end{equation*}
that is, $X$ is the Hamiltonian vectorfield with Hamiltonian the special K\"ahler potential. From the first equation we obtain:
\begin{align*}
0 &= (\partial_i J_k^j) X^k\\
&= (\partial_i g_{kr})\omega^{rj} X^k\\
&= \omega^{rj}(X^k\partial_k g_{ij})\\
&= -\omega^{jr}((L_X g)_{ji} + g([X, \partial_j], \partial_i) + g(\partial_j, [X, \partial_i]))\\
&= -\omega^{jr}((L_X g)_{ji} + \omega_{ji} + \omega_{ij})\\
&= -\omega^{jr}(L_X g)_{ji}.
\end{align*}
Therefore equation (\ref{homotheticvf1}) is equivalent to:
\begin{equation}
L_X g = 0. \label{killingX}
\end{equation}
Since $X$ determines $g$, the above is a differential equation for $K$ and defines a particular class of special K\"ahler metrics, namely the conical ones. Another way to read equation (\ref{killingX}) is as follows:
\begin{align*}
0 &= \omega^{kl}\partial_k K \partial_l g_{ij}\\
&= 2 g^{ks} \partial_k K C_{sij}.
\end{align*}
From this it follows:
\begin{equation*}
C_{ijk} H^k = 0.
\end{equation*}
We thus arrive at our best suited definition for conic special K\"ahler manifolds:
\begin{definition}
(\textbf{Conic special K\"ahler manifold - 2}) A conic special K\"ahler manifold is an affine special K\"ahler manifold whose symplectic vectorfield $X$, defined locally through $dK = \iota_X \omega$ is simultaneously a Killing vectorfield for the K\"ahler metric $g$.
\end{definition}
\noindent
The existence of the vectorfield $X$ implies that the function $K$ can be extended from a special Darboux patch to any simply connected patch containing it, in particular to a patch which is dense in $M$.

\medskip
\noindent
Now we shall shortly digress to recover the above geometric structure from the point of view of quantization. For this we will have to find yet two other ways to write identity (\ref{killingX}):
\begin{align*}
0 &= \omega^{kl}\partial_kK \partial_lg_{ij}\\
&= \omega^{kl}\partial_kK \partial_l \partial_i \partial_j K\\
&= \partial_i (\omega^{kl}g_{lj} \partial_k K) - \omega^{kl}g_{ik}g_{lj},
\end{align*}
which becomes:
\begin{equation*}
\partial_i (J_j^k \partial_k K) = \omega_{ij}
\end{equation*}
and because of the non-degeneracy of $\omega$ this is equivalent to:
\begin{equation*}
\partial_i(g^{jk} \partial_k K) = \delta_i^j,
\end{equation*}
which integrates to:
\begin{equation*}
g^{ik}\partial_k K = x^i,
\end{equation*}
from which:
\begin{align*}
\partial_k K &= g_{ki}x^i\\
&= \partial_k (\partial_i K x^i) - \partial_k K.
\end{align*}
Therefore, up to an irrelevant constant that can be absorbed in the definition of $K$:
\begin{equation*}
2K = \partial_i K x^i = \partial_i K g^{ij} \partial_j K = ||X||_g^2,
\end{equation*}
meaning in particular that $K$ describes a conic special K\"ahler manifold if and only if it is homogeneous of degree $2$ in special Darboux coordinates. 
\subsection{Digression: a guess for the quantum origin of the conic property}\label{quantumconic}
In this section we shall attempt a guess for the quantum origin of the conic property for a special K\"ahler manifold. It seems as though it comes from the requirement that the special K\"ahler potential $K$ be the classical counterpart of a quantum Hamiltonian $\hat{K}$ that preserves the space of coherent states. More precisely, $\hat{K}$ should preserve the space of coherent states as a subspace of $\mathcal{H}$, thus with no phase ambiguities, provided $\alpha_k$ is set to:
\begin{equation}
\alpha_k = - \frac{1}{2}\omega_{ik}x^i. \label{canonalphaconic}
\end{equation}
Thus we have the property:
\begin{equation*}
K(p) = \! _s\langle p| \hat{K} | p \rangle_s
\end{equation*}
and the statement that $\hat{K}$ preserves the space of coherent states with no phase ambiguities, is:
\begin{equation}
\exp(i \hat{K} t) |p \rangle_s = | \chi_t (p) \rangle_s, \label{conicevolution}
\end{equation}
where by $\chi_t$ we denote the canonical flow of $K$. Infinitesimally the above reads:
\begin{align}
i \hat{K} | p \rangle_s &= X^k A_k(p) |p \rangle_s \nonumber\\
&= -i X^k(\alpha_k + \omega_{ik}\hat{x}^i) | p \rangle_s. \label{conicevolutioninf}
\end{align}
Applying $\!_s\langle p|$ to the above equation we obtain the desired result:
\begin{align*}
K &= \frac{1}{2}X^k \omega_{ik}x^i - \!_s \langle p | \hat{x}^i | p \rangle_s \omega_{ik}X^k\\
&= -\frac{1}{2}\partial_i K x^i + \partial_i K \; \!_s \langle p | \hat{x}^i | p \rangle_s\\
&= \frac{1}{2} \partial_i K x^i.
\end{align*}
In the last step we used the fact that $\!_s \langle p | \hat{x}^i | p \rangle_s  = x^i$. This follows from:
\begin{align*}
\partial_j (\; \!_s \langle p | \hat{x}^i | p \rangle_s)  &= \!_s \langle p | [A_j(p), \hat{x}^i] | p \rangle_s\\
&= i\omega_{kj}[\hat{x}^k, \hat{x}^i]\\
&= \delta_j^i.
\end{align*}
Now we shall ask under what changes of gauge for $\alpha$, equation (\ref{conicevolutioninf}) remains unaltered\footnote{Notice that for a conic special K\"ahler manifold, the gauge (\ref{canonalphaconic}) is the same as canonical gauge (\ref{canonalpha}).}. Clearly the gauge transformations are reduced to:
\begin{equation*}
|p \rangle_s \rightarrow \exp(if(p))|p \rangle_s \s\s\s \textrm{with } f \in C^{\infty}(M)  \s\s\s \textrm{such that } X \cdot f = 0.
\end{equation*}
Now, the structure of a conic special K\"ahler manifold is necessary for (\ref{conicevolution}) to be fulfilled, but it is by no means sufficient. Indeed, in the simplest case of $\mathbb{R}^{2d}$ one can check that the admissible $K$'s are reduced to quadratic ones. What distinguishes the case $M = \mathbb{R}^{2d}$ is not the connection $A$, but rather an initial choice $|p_0 \rangle_s$ for a definite marked point $p_0 \in M$. Thus turning the argument around, once $K$ is fixed, equation (\ref{conicevolution}) puts constraints on this choice. Let
\begin{equation*}
\Delta(p_2, p_1) := \!_s \langle p_2 | p_1 \rangle_s, 
\end{equation*}
then, one such natural constraint is:
\begin{equation*}
\Delta(p_2, p_1) = \Delta (\chi_t(p_2), \chi_t(p_1)).
\end{equation*}
Infinitesimally, the above becomes the following equation:
\begin{equation*}
-i (\partial_i K(p_2) - \partial_i K(p_1))\omega^{ik}\frac{\partial}{\partial x_1^k} \Delta(p_2, p_1) = \frac{1}{2}\partial_i K(p_2) (x_2^i - x_1^i)\Delta(p_2, p_1).
\end{equation*}
As stated before, for fixed $K$ this can be viewed as a differential equation for $\Delta$ while, for fixed $\Delta$ it can be viewed as a constraint on the choice of $K$. There is in fact an even more elementary constraint on $K$ if we allow $M$ to contain the point at the origin of the coordinate system. Then, indeed, the only homogeneous degree 2 functions are the quadratic ones. Therefore, in the more general case we need to assume that $0 \notin M$. This condition however is but a consequence of the further requirement entailed in definition \ref{defconic1} that the action of $X$ be free as $0$ would clearly be a fixed point. In fact it is the unique fixed point and there the metric is singular.

\subsection{Coping with negative signature}\label{negsig}
In this section we shall investigate how the definition of quantization should be modified in the case of a non-Riemannian K\"ahler manifold. This is of interest since precisely moduli spaces of $N = (2,2)$ 2-d super conformal field theories are of this type. Recall that in the discussion of the coherent tangent bundle it was crucial that $M$ be Riemannian, otherwise the definition of coherent state would have implied the existence of negative normed states in the Hilbertspace. Let's start with the simplest case, namely again $M = \mathbb{R}^2$, this time however we shall change the K\"ahler structure as follows: $\omega \rightarrow - \omega$, $J \rightarrow J$ and $g \rightarrow -g$. Under this change, the Heisenberg algebra is changed to:
\begin{equation*}
[\hat{x}, \hat{p}] = -i
\end{equation*}
and in terms of the annihilation operator $a = (1/\sqrt{2})(\hat{x} + i\hat{p})$ and creation operator $b = (1/\sqrt{2})(\hat{x} - i\hat{p})$:
\begin{equation*}
[a, b] = -1.
\end{equation*}
We shall now define the highest weight state $v_0$ as before through:
\begin{equation*}
a v_0 = 0.
\end{equation*}
Then a basis for the highest weight representation is furnished by:
\begin{equation*}
v_n = \frac{b^n}{\sqrt{n!}}v_0, \s n \in \mathbb{N}_0.
\end{equation*}
At this point we realize the impossibility of finding, in this representation, a positive definite sesquilinear bilinear form with respect to which $\hat{x}$ and $\hat{p}$ are self-adjoint. Indeed given such a bilinear form $B$, we would have:
\begin{equation*}
0 < B(b v_0, b v_0) = B(v_0, a b v_0) = B(v_0, [a, b]v_0) = - B(v_0, v_0),
\end{equation*}
which is clearly contradictory. Instead, what we can require is the existence of two bilinear forms $B_+$ and $B_-$. We shall require the former to be positive definite and sesquilinear, thus introducing a Hilbertspace topology on the representation. However $B_+$ will have the property that $\hat{x}$ and $\hat{p}$ are anti-self-adjoint with respect to it. On the other hand $B_-$ is non-degenerate sesquilinear such that $\hat{x}$ and $\hat{p}$ are self-adjoint, but it will be indefinite. Under the normalization $B_+(v_0, v_0) = B_-(v_0, v_0) = 1$ we thus obtain:
\begin{equation*}
B_+ (v_m, v_n) = \delta_{m, n} \s\s\s B_-(v_m, v_n) = (-1)^n \delta_{m, n}.
\end{equation*}
Thus $\{v_n\}_{n \in \mathbb{N}_0}$ form an orthonormal basis of the Hilbertspace $\mathcal{H}$. We can define $B_-$ in terms of $B_+$ as:
\begin{equation*}
B_-(\,\cdot\,, \,\cdot\,) = B_+(\,\cdot\, , (-1)^{ba} \cdot).
\end{equation*}
A suggestive way of interpreting $(-1)^{ba}$ is as $(-1)^F$ where $F$ is a ``fermion-number" operator. For this we need to introduce the following decomposition of $\mathcal{H}$:
\begin{equation*}
\mathcal{H} = \mathcal{H}_{even} \oplus \mathcal{H}_{odd},
\end{equation*}
where $\mathcal{H}_{even}$ is the subspace spanned by $v_n$ with $n$ even and $\mathcal{H}_{odd}$ is defined analogously. Then we define fermion fields $\psi_1$ and $\psi_2$ by:
\begin{align*}
\psi_1 v_{2n} &= 0 \s\s\s\s\s\s\; \psi_1 v_{2n + 1} = v_{2n} \s \\
\psi_2 v_{2n} &= v_{2n + 1} \s\s\s \psi_2 v_{2n + 1} = 0 \s \forall n \in \mathbb{N}_0.
\end{align*}
Then $\psi_1$ and $\psi_2$ satisfy the following properties:
\begin{equation*}
\{\psi_1, \psi_2\} = 1
\end{equation*}
and
\begin{equation*}
B_+(\psi_1 \cdot, \cdot) = B_+(\cdot, \psi_2 \cdot).
\end{equation*}
Finally we can define $F$ by:
\begin{equation*}
F =\psi_2 \psi_1.
\end{equation*}
On $\mathcal{H}$ we can furthermore define the differential:
\begin{equation*}
Q = \psi_1 ba.
\end{equation*}
This allows us to single out $v_0$, which is the unique invariant state under the $U(1)$ action:
\begin{equation*}
R(\theta) := e^{i\theta ba},
\end{equation*}
as the unique representative of $Q$-cohomology $H_Q(\mathcal{H})$.

\medskip
\noindent
Now let's turn to coherent states. These are now given by:
\begin{equation*}
|\alpha \rangle = \exp( - \alpha b + \overline{\alpha} a) v_0 = \exp\K{(}{)}{\frac{|\alpha|^2}{2}} \sum_{n \geq 0} \frac{(-\alpha)^n}{\sqrt{n!}}v_n.
\end{equation*}
Clearly the translation operator is not unitary any more, that is, it is not an isometry with respect to $B_+$, but it is an isometry with respect to $B_-$. It is in fact unbounded, but clearly its domain includes the coherent states. Now we shall generalize the above discussion to understand the structure of the coherent tangent bundle for an arbitrary Lorentzian K\"ahler manifold.
\subsection{The coherent tangent bundle in the Lorentzian case}\label{LCTB}
In this section we will construct the coherent tangent bundle in the case of a general Lorentzian K\"ahler manifold. An important result will be that contrary to the Riemannian case, in the present case the Hilbert-bundle is not necessarily trivial, in particular it cannot in general be trivialized on an entire Darboux patch. The caveat stems from the fact that in the Lorenztian case one needs to make a choice of two ``negative" directions, and this choice depends non trivially, not only on the symplectic form, but also on the metric, which contrary to the symplectic form, cannot be flat on an entire patch unless the Riemann curvature vanishes. In this section we will show however that if one makes a choice of negative directions at a given point $p_0 \in M$, this choice can be extended to an open neighborhood $V_+$ containing $p_0$. Ultimately the detailed choice at $p_0$ will be irrelevant.

\medskip
\noindent
We start by considering a point $p_0 \in M$ and erect a Darboux coordinate system in the neighborhood of $p_0$ such that at $p_0$ the metric is the standard Lorentzian metric\footnote{This can obviously also be achieved for special Darboux patches in special K\"ahler manifolds.}. The coherent tangent bundle at $p_0$ will then be the collection of states defined by:
\begin{equation*}
\hat{x}^T (\eta + i\epsilon) | u \rangle_{p_0} = u^T (\eta + i\epsilon) |u \rangle_{p_0}.
\end{equation*}
In particular there will be a state $|0 \rangle_{p_0}$. The representation of the Heisenberg algebra thus obtained with highest weight $|0 \rangle_{p_0}$, as we have observed in the above section, is naturally not a Hilbert space, but rather a vector space equipped with the pairing $B_-$, with respect to which $\hat{x}^i$ are hermitian, which is defined precisely as in section \ref{negsig} with $a$ and $b$ corresponding to the first coordinates $x^1$ and $x^{d+1}$. We also observed in the previous section that we can however endow this vectorspace with the structure of a Hilbert space $\mathcal{H}$ with scalar product $B_+$ with respect to which $\hat{x}^1$ and $\hat{x}^{d+1}$ are anti-hermitian while the rest are hermitian.

\medskip
\noindent
Now we ask how large the Darboux neighborhood $V_+$ of $p_0$ is allowed to be for the coherent states $|u \rangle_p$ for $p \in V_+$ to belong to the same Hilbert space $\mathcal{H}$. We shall denote by $Sp^{\epsilon}$ the symplectic group with symplectic form $\epsilon$, then the previous question is clearly equivalent to determining the subset:
\begin{equation*}
Sp^{\epsilon}_+ := \{ \Lambda \in Sp^{\epsilon} \; | \; |0 \rangle^{\Lambda} \in \mathcal{H}\},
\end{equation*}
where $|0 \rangle^{\Lambda}$ is defined through the condition:
\begin{equation}
(\Lambda^T \hat{x})^T (\eta + i\epsilon) | 0 \rangle^{\Lambda} = 0. \label{lambdavacuum}
\end{equation}
Then, denoting by $V^{max}$ the maximal Darboux patch containing $p_0$:
\begin{equation*}
V_+^{max} = \{ p \in V^{max} \; | \exists \Lambda \in Sp^{\epsilon}_+ \s s.t. \s g(p) = \Lambda \eta \Lambda^T\}.
\end{equation*}
As we have seen explicitly in section \ref{CTB}, in the Riemannian case, $Sp_+ = Sp$ and therefore $V_+^{max} = V^{max}$. This is but a consequence of the Stone von Neumann theorem that asserts, in particular, the uniqueness of unitary irreducible representations of the Heisenberg algebra. The Lorentzian case, however, corresponds to non-unitary representations, and indeed, as we will show, $Sp^{\epsilon}_+ \subsetneq Sp^{\epsilon}$. We will show however that $Sp_+^{\epsilon}$ contains an open neighborhood of the identity, a requirement to, at least locally, quantize $M$. As a concrete representation for $\mathcal{H}$ we choose the usual $L^2(\mathbb{R}^d)$ where $\hat{x}$ acts as:
\begin{equation*}
\hat{x} = \tilde{E}\K{(}{)}{\begin{array}{cc} q \\ -i \nabla_q \end{array}},
\end{equation*}
where:
\begin{equation*}
\tilde{E} := \K{(}{)}{\begin{array}{cc} E & 0 \\ 0 & E \end{array}}
\end{equation*}
and:
\begin{equation*}
E = \K{(}{)}{\begin{array}{cc} i & 0 \\ 0 & 1_{(d-1 \times d-1)} \end{array}}.
\end{equation*}
Then, resorting to the notation in section \ref{sol2}, equation (\ref{lambdavacuum}) becomes:
\begin{align*}
\K{(}{)}{\begin{array}{cc} R_1 & R_2 + iE^2 \\ R_2^T - iE^2 & R_4 \end{array}} \tilde{E}\K{(}{)}{\begin{array}{cc} q \\ -i \nabla_q \end{array}} \psi_0(q, \Lambda) = 0,
\end{align*}
with $\psi_0(q, \Lambda)$ the wavefunction corresponding to $|0\rangle^{\Lambda}$. The above is equally well written as:
\begin{equation*}
\nabla_q \psi = i\tilde{\tau} q \psi,
\end{equation*}
where:
\begin{equation*}
\tilde{\tau} := - E^{-1}(R_2 + iE^2)^{-1} R_1E = -E^{-1}R_4^{-1} (R_2^T - iE^2)E,
\end{equation*}
thus:
\begin{equation*}
\psi(q) = \mathcal{N}\exp(i \langle q, \tilde{\tau} q \rangle)
\end{equation*}
and $\psi \in L^2(\mathbb{R}^d)$ provided:
\begin{equation}
\mathrm{Im}\,\tilde{\tau} > 0. \label{normalizabilitycond}
\end{equation}
Since the above is an open set, by local continuity of $\Lambda(p)$, $g(\Lambda)$ and $\tilde{\tau}(g)$, $V_+^{max}$ contains an open neighborhood of $p_0$. One can easily check by way of counterexample that condition (\ref{normalizabilitycond}) is non trivial and in particular  $Sp^{\epsilon}_+ \subsetneq Sp^{\epsilon}$.

\medskip
\noindent
Consider now the complex modulus:
\begin{equation*}
\tau := E \tilde{\tau} E^{-1}.
\end{equation*}
Clearly $\tau$ is in the Siegel upper half space. From this we deduce the complete characterization of $Sp^{\epsilon}_+$:
\begin{equation*}
Sp^{\epsilon}_+ = \{\Lambda \in Sp^{\epsilon} \; | \mathrm{Im} ( E^{-1}\tau(\Lambda) E) > 0\}.
\end{equation*}
At this point we shall study the states $|u \rangle^{\Lambda}$, which we shall rename $|u \rangle_{\tau}$. Following the analogous steps for $|0\rangle^\Lambda$ we obtain that the corresponding wavefunction $\psi_u (\tau, q)$ is of the form:
\begin{equation*}
\psi_u (\tau, q) = \mathcal{N}(\tau, u) \exp\K{(}{)}{ \frac{i}{2} \langle (q - E^{-1}u_q), \tilde{\tau} (q - E^{-1}u_q) \rangle + i \langle E^{-1} u_p, q \rangle}.
\end{equation*}
Just as the normalization constant in section \ref{sol2}, here $\mathcal{N}(\tau, u)$ is fixed by three analogous conditions. The first is a normalization condition with respect to $B_-$ instead of $B_+$:
\begin{equation*}
1 = \!_{\tau}\langle u | (-1)^F |u \rangle_{\tau} = \int_{\mathbb{R}^d} d^dq\;\overline{\psi}_u(\tau, q) \psi_u(\tau, E^2 q).
\end{equation*}
Solving for $|\mathcal{N}|$ yields:
\begin{equation*}
|\mathcal{N}(\tau, u)| = \pi^{-d/4}(\det \mathrm{Im}\,\tau)^{1/4}.
\end{equation*}
The second condition on $\psi_u(\tau, q)$ is equation (\ref{defup2}) which remains unchanged in the Lorentzian case. As in section \ref{sol2}, let $\theta$ be defined through:
\begin{equation*}
\mathcal{N}(\tau, u) = |\mathcal{N}(\tau)|e^{i\theta(\tau, u)}.
\end{equation*}
Then (\ref{defup2}) is equivalent to:
\begin{equation*}
\nabla_{u_q}\theta + E^2 \overline{\tau} E^2 \nabla_{u_p} \theta = -\frac{1}{2} E^2 \overline{\tau} u_q - \frac{1}{2} E^2 u_p,
\end{equation*}
the solution to which is:
\begin{equation*}
\theta(\tau, u_q, u_p) = -\frac{1}{2}\langle u_p, E^2 u_q\rangle + \gamma(\tau).
\end{equation*}
Therefore in the Lorentzian case, the wavefunction of the coherent state $|u \rangle_{\tau}$ is given by:
\begin{align*}
\psi_u(\tau, q) &= \pi^{-d/4}(\det \mathrm{Im}\,\tau)^{1/4} \exp (i \gamma(\tau)) \cdot \\
&\exp\K{(}{)}{ \frac{i}{2} \langle (q - E^{-1}u_q), \tilde{\tau} (q - E^{-1}u_q) \rangle + i \langle E^{-1} u_p, q \rangle - \frac{i}{2}\langle u_p, E^2 u_q \rangle}.
\end{align*}
Using (\ref{defup3}) we can again fix the phase. For $M$ affine special K\"ahler, on a special Darboux patch we obtain:
\begin{equation*}
\partial_k \gamma = -\beta_k - \frac{1}{2} \mathrm{tr} ([(\Gamma_k \omega)_{pp}] \,\mathrm{Im} \,\tau \; E^2).
\end{equation*}
At this point we can compute the overlap $\!_{\tau_2}\langle u_2 | (-1)^F |u_1 \rangle_{\tau_1}$, which, as in the Riemannian case, essentially corresponds to the propagator of the master equation:
\begin{align*}
&\! _{\tau_2}\langle u_2 | (-1)^F|u_1 \rangle_{\tau_1} := \int_{\mathbb{R}^n} \overline{\psi_{u_2}}(\tau_2, q) \psi_{u_1}(\tau_1, E^2 q)\\
&=(2i)^{d/2}\frac{(\det \mathrm{Im} \tau_1)^{1/4} (\det \mathrm{Im} \tau_2)^{1/4}}{(\det (\tau_1 - \overline{\tau_2}))^{1/2}} \cdot \\
&\exp \K{(}{)}{ - \frac{i}{2}\langle E^2u_{q,1}, z_1 \rangle + \frac{i}{2} \langle E^2 u_{q,2}, \overline{z}_2\rangle - \frac{i}{2} \langle E^2(z_1 - \overline{z}_2), (\tau_1 - \overline{\tau}_2)^{-1} (z_1 - \overline{z}_2)\rangle}\cdot\\
& \exp\K{(}{)}{i \int_{p_0}^{p} \K{(}{)}{\beta + \frac{1}{2} \mathrm{tr} ((\Gamma_k \omega)_{pp} \mathrm{Im} \tau \, E^2) dx^k}},
\end{align*}
where, as in the Riemannian case, we have introduced the complex coordinates $z = u_p - \tau u_q$. Thus the expression is identical to the one in the Riemannian case with the only difference that the bilinear form on configuration space is now the standard Minkowski bilinear form $\langle E^2 \,\cdot\,, \,\cdot\,\rangle$ instead of the standard scalar product. Anlogously to the case of Riemannian affine special K\"ahler manifolds where the propagator is given by (\ref{propagator1}), we shall see in the next section that in the Lorentzian case the propagator is given by:
\begin{align*}
&K(u, p, x + y, p_0) := \!_{p, A_s} \langle u| (-1)^F | p \rangle_s^y\\
&=(2i)^{d/2}\frac{(\det \mathrm{Im} \tau_1)^{1/4} (\det \mathrm{Im} \tau_2)^{1/4}}{(\det (\tau_1 - \overline{\tau_2}))^{1/2}} \cdot \\
& \exp\K{(}{)}{-i \int_{p_0}^{p} \K{(}{)}{(\alpha - \beta) - \frac{1}{2}\omega_{kj}(x + y)^j dx^k - \frac{1}{2} \mathrm{tr} ((\Gamma_k \omega)_{pp} \mathrm{Im} \tau \, E^2) dx^k}}\cdot\\
&\exp \K{(}{)}{ - \frac{1}{4}||u||_{g(p)}^2}\cdot\\
&\exp\K{(}{)}{ - \frac{i}{2}\langle E^2 u_{q,1}, z_1 \rangle + \frac{1}{4} \langle \overline{z}_2, R_4(p) \overline{z}_2\rangle - \frac{i}{2} \langle E^2(z_1 - \overline{z}_2), (\tau_1 - \overline{\tau}_2)^{-1} (z_1 - \overline{z}_2)\rangle},
\end{align*}
where we have used the same notation as in section \ref{sol2}.
\subsection{Remarks on the quantization of Lorentzian conic special K\"ahler manifolds}\label{lorentzquantum}
In this section we shall first discuss how the quantization of Riemannian affine special K\"ahler manifolds translates to the Lorentzian case, show how to project to positive normed states, and then discuss normalization conditions of the wavefunction $Z(u,p)$ thus presenting the form of the general solution to the master equation. We shall develop the first point in the form of a series of remarks:
\begin{itemize}
\item To quantize an appropriate Darboux neighborhood ($V_+$) of $p_0 \in M$, one chooses the Darboux coordinates such that $g(p_0) = \eta$.
\item Locally quantization involves a triple $(V_+, \phi, \mathbb{P}^{\infty})$, but contrary to the Riemannian case, now $\mathbb{P}^{\infty}$ is endowed with the pairing
\begin{equation*}
(v, w)_- = \frac{|B_-(v, w)|}{\K{(}{)}{B_-(v,v)B_-(w,w)}^{1/2}}
\end{equation*}
and the group of automorphisms of $\mathbb{P}^{\infty}$ is defined accordingly.
\item The flat connection $A$ does thus no longer induce a unitary parallel transport, but rather a parallel transport that is an isometry w.r.t. $B_-$.
\item One must choose generators of the Heisenberg algebra $\hat{x}^i$, such that $\hat{x}^1$ and $\hat{x}^{d + 1}$ are anti-hermitian w.r.t. $B_+$, while the rest are hermitian.

\item The form of the operator $S_{\Sigma}$ introduced as $S$ in (\ref{SofSigma}) is left unchanged, and it is now an isometry w.r.t. $B_-$.

\item As a consequence in order for the tensorial property (\ref{Zastensor}) of $Z(u,p)$ to hold, the definition of the wavefunction must be replaced by:
\begin{equation*}
Z(u, p) := \, \!_{p, A}\langle u| (-1)^F | p \rangle_A.
\end{equation*}
\end{itemize}
With the above modifications the quantization procedure of affine Lorentzian special K\"ahler manifolds proceeds without change as the one for Riemannian affine special K\"ahler manifolds until the end of section \ref{MEsection} with the only exception of the normalization conditions (\ref{norm1}, \ref{norm2}, \ref{norm3}). One last remark regards section \ref{quantumconic} where, in the Lorentzian case, all matrix elements of the form $\langle p_2| \mathcal{O} |p_1\rangle$ must be replaced with $\langle p_2| (-1)^F \mathcal{O} |p_1\rangle$.
\subsubsection{Projecting onto ``positive normed" states: the coherent horizontal bundle}
Let $M$ be a conic special K\"ahler manifold of dimension $2d$, we shall distinguish between three regions of $M$:
\begin{align*}
M_{+} &:= \{p \in M | K(p) > 0\},\\
M_0 &:= \{p \in M | K(p) = 0\},\\
M_- &:= \{p \in M | K(p) < 0\}.
\end{align*}
As discussed earlier $M_0$ is singular with a conic singularity approaching $x = 0$. Now we shall concentrate on $M_-$. There, an orthonormal basis of negative or ``timelike directions" in the tangent bundle $TM_-$ is given by the hamiltonian vectorfield $X$ and $JX$. Indeed:
\begin{equation*}
g(X, X) = g(JX, JX) = 2K < 0.
\end{equation*}
Therefore, on the orthogonal complement with respect to $g$ of $X$ and $JX$, $g$ is positive definite. We thus define the horizontal bundle as:
\begin{equation*}
HM := \{V \in TM_- | \; g(V, X) = g(V, JX) = 0\}.
\end{equation*}
In particular $HM$ is the image of a section $P \in \Gamma(M_- , \mathrm{End}(TM_-))$ of projections $P(p)$, which in special coordinates is given by:
\begin{equation}
P_k^j = \delta_k^j - \frac{1}{2}\partial_k \log |K| g^{ij} \partial_i K + \frac{1}{2} J_k^l \partial_l \log |K| \omega^{ij} \partial_iK.\label{projection}
\end{equation}
Corresponding to $HM$ there is a quantum counterpart that we shall name coherent horizontal bundle, defined as the sub-bundle of the trivial Hilbert-bundle, given by the image of the section of projection operators $\mathcal{P} \in \Gamma(M_- , \mathrm{End}(\mathcal{H}))$, where $\mathcal{P}(p, A)$ is an orthogonal projection at every point $p$. This projection is the obvious generalization of the projector onto $v_0$ of section [\ref{negsig}]. Thus, the action of $\mathcal{P}(p, A_s)$ on the basis $|u \rangle_{p, s}$ is given by:
\begin{align*}
\mathcal{P}(p, A_s) | u \rangle_{p, s} &:= \exp\K{(}{)}{- \frac{1}{4}||(1 - P(p))u||_{g(p)}^2} | P(p) u \rangle_{p,s}\\
&= \exp\K{(}{)}{-\frac{1}{2K(p)} |g(H(p), u)|^2} | P(p) u \rangle_{p,s}.
\end{align*}
It is an easy exercise to check that $\mathcal{P}$ is self-adjoint w.r.t $B_-$.
\subsubsection{Normalization conditions and the general solution}
Now we shall pass to normalization conditions. In the Lorentzian case, equations (\ref{norm2}, \ref{norm3}) are modified to:
\begin{align*}
(-1)^F = \frac{1}{(2\pi)^n} \int_{T_pM_-}&\, du^1 \wedge \cdots \wedge du^{2d} \sqrt{\mathrm{det} \, g}\cdot \\
&\exp\K{(}{)}{\frac{1}{K(p)} |g(H(p), u)|^2}| Pu - (1-P)u \rangle_{p, A} \;\! _{p, A}\langle  u |.
\end{align*}
And the normalization is with respect to $B_-$ rather than $B_+$, therefore:
\begin{align*}
1 = \frac{1}{(2\pi)^n} \int_{T_pM_-} &\sqrt{\mathrm{det} \, g} \, \exp\K{(}{)}{\frac{1}{K(p)} |g(H(p), u)|^2} \cdot \\ 
&\overline{Z_A}( Pu - (1 - P)u, p)Z_A(u, p) \, du^1 \wedge \cdots \wedge du^{2d}.
\end{align*}
It follows that the general solution to the master equation is then given by:
\begin{align}
Z(u, p)^f = \int_{T_{p_0}M} &dy^1 \wedge \cdots \wedge dy^{2d}\sqrt{\det g_s(p_0)}  \exp\K{(}{)}{\frac{1}{K(p_0)} |g(H(p_0), y)|^2} \cdot \nonumber\\
&K_-(u, p, x_s + y, p_0) \exp\K{(}{)}{-\frac{1}{4}||y||_{g_s(p_0)}^2}f((1 + iJ_0)y),\label{lorentzsol}
\end{align}
where $f$ is an arbitrary normalizable function w.r.t. $B_-$.
\subsection{The quantization of projective special K\"ahler manifolds}\label{projsk}
In this section we will construct the wavefunction $Z_{red,A}(u, p)$ for an arbitrary projective special K\"ahler manifold $\tilde{M}$ of dimension $2d$ that arises as a holomorphic quotient of a Lorentzian conic special K\"ahler manifold $M$ of dimension $2d + 2$. First of all, it is convenient at this point to introduce complex coordinates and express $H$ in terms of these. We shall stay in the special Darboux coordinate system, and erect corresponding holomorphic coordinates $(z^0, \dots, z^d)$. Then (\ref{homotheticvf}) becomes:
\begin{equation*}
\partial_{\mu} H^{\nu} = \delta_{\mu}^{\nu}.
\end{equation*}
Therefore:
\begin{equation*}
H = z^{\mu}\partial_{\mu},
\end{equation*}
where the vector of complex special coordinates is related to the vector of Darboux coordinates $x = (x_q, x_p)$ via:
\begin{equation*}
z = x_p - \tau x_q.
\end{equation*}
The quotient of $M$ by $H$ clearly has as holomorphic functions the ones defined on $M$ of homogeneous degree $0$, therefore $\tilde{M}$ can be covered by affine patches as $\tilde{M}_0$ with coordinates $(y^1, \dots, y^d)$ given by:
\begin{equation*}
(z^0, z^1, \dots, z^{d}) =: (\lambda, \lambda y^1, \dots, \lambda y^d),
\end{equation*}
with $\lambda \neq 0$. In this new coordinate system $(\lambda, y^1, \dots, y^d)$:
\begin{equation*}
H = \lambda \frac{\partial}{\partial \lambda}.
\end{equation*}
From now on we shall label the coordinates $y$ and $z$ with ($\alpha, \beta, \gamma$) and ($\mu, \nu, \rho, \sigma$) respectively. Analogously we will label the corresponding real coordinates with non-capital and capital latin letters respectively. At this point we can express the projection $P$ introduced in (\ref{projection}) in complex coordinates:
\begin{align*}
P_{\mu}^{\nu} &= \delta_{\mu}^{\nu} - \frac{1}{2} \partial_{\mu} \log |K| g^{\nu \overline{\rho}} \partial_{\overline{\rho}} K + \frac{1}{2} J_{\mu}^{\sigma} \partial_{\sigma} \log |K| \omega^{\nu \overline{\rho}} \partial_{\overline{\rho}} K\\
&= \delta_{\mu}^{\nu} - \frac{1}{2} \partial_{\mu} \log |K| (g^{\nu \overline{\rho}} - i\omega^{\nu \overline{\rho}})\partial_{\overline{\rho}} K \\
&=  \delta_{\mu}^{\nu} - \partial_{\mu} \log |K| g^{\nu \overline{\rho}}\partial_{\overline{\rho}} K\\
&= \delta_{\mu}^{\nu} - z^{\nu}\partial_{\mu} \log |K| \\
P_{\overline{\mu}}^{\nu} &= 0\\
P_{\mu}^{\overline{\nu}} &= 0\\
P_{\overline{\mu}}^{\overline{\nu}}&=  \delta_{\overline{\mu}}^{\overline{\nu}} - z^{\overline{\nu}}\partial_{\overline{\mu}} \log |K|.
\end{align*}
Therefore, in particular, in special coordinates we have the following holomorphic frame for the horizontal bundle:
\begin{align*}
V_{\alpha} &= \Sigma_{\alpha}^{\mu} P_{\mu}^{\nu} \partial_{\nu}\\
&= \frac{\partial z^{\mu}}{\partial y^{\alpha}} (\delta_{\mu}^{\nu} - z^{\nu}\partial_{\mu} \log |K|)\partial_{\nu}\\
&= \frac{\partial}{\partial y^{\alpha}} - \K{(}{)}{\frac{\partial}{\partial y^{\alpha}} \log |K|} z^{\nu}\partial_{\nu}\\
&= \frac{\partial}{\partial y^{\alpha}} - \K{(}{)}{\frac{\partial}{\partial y^{\alpha}} \log |K|} \lambda \frac{\partial}{\partial \lambda},
\end{align*}
where $\Sigma_{\alpha}^{\mu} = dz^{\mu}/dy^{\alpha}$. We can now define the wavefunction reduced to the projective special K\"ahler manifold:
\begin{definition}
The quantization of the holomorphic quotient $\tilde{M}$ is given by the reduced wavefunction:
\begin{equation*}
Z_{red, A}(u, p) := \! _ {p, A}\langle \Sigma^T u | \mathcal{P}^{\dagger}(p, A) (-1)^F | p \rangle_A.
\end{equation*}
\end{definition}
Therefore:
\begin{align*}
Z_{red, A}(u,p) := \exp \K{(}{)}{-\frac{1}{4}||P(p)\Sigma^T u||^2_{g(p)} - \frac{1}{2K(p)}|g(H(p), \Sigma^T u)|^2}\mathcal{C}_{red}(u^i \partial_i),
\end{align*}
where
\begin{align}
\mathcal{C}_{red} = \iota^* (\mathcal{C}\circ P),\label{Cred}
\end{align}
and by $\iota$ we have denoted the inclusion of the level set $\lambda$ in $M$. In particular we have
\begin{equation*}
\mathcal{C}_{red} = \exp\K{(}{)}{\sum_{n \geq 0} \frac{(-1)^n}{n!}\mathcal{C}_{red}^n},
\end{equation*}
with, in special coordinates:
\begin{equation*}
(\mathcal{C}_{red}^n)_{\overline{\alpha}_1, \dots, \overline{\alpha}_n} = \mathcal{C}^n_{\overline{\mu}_1, \dots, \overline{\mu}_n} \K{(}{)}{\frac{\partial z^{\overline{\mu}_1}}{\partial y^{\overline{\alpha}_1}} - z^{\overline{\mu}_1} \frac{\partial}{\partial y^{\overline{\alpha}_1}} \log |K|} \cdots \K{(}{)}{\frac{\partial z^{\overline{\mu}_n}}{\partial y^{\overline{\alpha}_n}} - z^{\overline{\mu}_n} \frac{\partial}{\partial y^{\overline{\alpha}_n}} \log |K|}
\end{equation*}
and:
\begin{align*}
\frac{\partial z^{0}}{\partial y^{\alpha}} - z^{0} \frac{\partial}{\partial y^{\alpha}} \log |K| &= - \lambda \frac{\partial}{\partial y^{\alpha}} \log |K|\\
\frac{\partial z^{\beta}}{\partial y^{\alpha}} - z^{\beta} \frac{\partial}{\partial y^{\alpha}} \log |K| &= \lambda \K{(}{)}{\delta_{\alpha}^{\beta} - y^{\beta} \frac{\partial}{\partial y^{\alpha}} \log |K|}.
\end{align*}
We shall extend the $y$ coordinate system to incorporate $y^0 := 1$, and define $h$ through:
\begin{equation*}
K(z, \overline{z}) = -|\lambda|^2 h(y, \overline{y}),
\end{equation*}
then:
\begin{equation*}
(\mathcal{C}_{red}^n)_{\overline{\alpha}_1, \dots, \overline{\alpha}_n} = \overline{\lambda}^n \, \mathcal{C}^n_{\overline{\mu}_1, \dots, \overline{\mu}_n} \K{(}{)}{\delta_{\overline{\alpha}_1}^{\overline{\mu}_1} - y^{\overline{\mu}_1} \frac{\partial}{\partial y^{\overline{\alpha}_1}} \log h} \cdots \K{(}{)}{\delta_{\overline{\alpha}_n}^{\overline{\mu}_n} - y^{\overline{\mu}_n} \frac{\partial}{\partial y^{\overline{\alpha}_n}} \log h}.
\end{equation*}
At this stage we can determine the master equation satisfied by $\mathcal{C}_{red}$. We shall proceed analogously to the affine case. Thus we start by collecting the following computational building blocks. The first crucial building block is the K\"ahler structure on the projective manifold $\tilde{M}$:
\begin{align}
\tilde{g}_{\alpha \overline{\beta}} &= \Sigma_{\alpha}^{\mu} P_{\mu} ^{\rho} g_{\rho \overline{\sigma}} P_{\overline{\nu}}^{\overline{\sigma}}\Sigma_{\overline{\beta}}^{\overline{\nu}} = \overline{\lambda} \Sigma_{\alpha}^{\mu} P_{\mu}^{\rho} g_{\rho \overline{\beta}} \nonumber \\
&= \Sigma_{\alpha}^{\mu} \K{(}{)}{g_{\mu \overline{\nu}} - \frac{1}{K} \partial_{\mu} K \partial_{\overline{\nu}} K}\Sigma_{\overline{\beta}}^{\overline{\nu}} \nonumber \\
&= -|\lambda|^2 \K{(}{)}{\partial_{\alpha}\partial_{\overline{\beta}}h - \frac{1}{h} \partial_{\alpha} h \partial_{\overline{\beta}}h} \nonumber\\
&= K \partial_{\alpha}\partial_{\overline{\beta}} \log h. \label{planckrescaling}
\end{align}
The form obtained in the last step shows that $\tilde{g}$ is indeed a K\"ahler metric, not on the holomorphic quotient of $M$ by the action of $H$, but rather on the symplectic quotient of $M$ by the action of $X$ where $K$ is constant. Indeed the above precisely defines the Marsden-Weinstein quotient. We thus define the normalized K\"ahler metric:
\begin{equation*}
\hat{g}_{\alpha \overline{\beta}} := -\partial_{\alpha} \partial_{\overline{\beta}} \log h.
\end{equation*}
We shortly digress to observe that formula (\ref{planckrescaling}) means that the value of $K$ on the corresponding symplectic quotient is related to Planck's constant via:
\begin{equation*}
K = -\frac{1}{\hbar}.
\end{equation*}
In other words, Planck's constant precisely labels the choice of symplectic quotient:
\begin{equation*}
M_{\hbar} \sim K^{-1}(-\hbar^{-1})/S^1.
\end{equation*}
Here $\sim$ means homeomorphic.

\medskip
\noindent
Now we shall consider the dependence of $\mathcal{P}(p, A) | \Sigma^T u \rangle_{p, A}$ on $p$. We shall do this in steps. First we shall consider the dependence on $p$ of the canonical coherent state $|\Lambda^T P \Sigma^T u \rangle$, where we have used the same notation as in section \ref{CTB}. We obtain:
\begin{align}
\partial_K |\Lambda^T P\Sigma^T u \rangle &= \langle \partial_K (\Lambda^T P \Sigma^T)u, \Lambda^{-1} \nabla_{P\Sigma^T u} \rangle | \Lambda^T P \Sigma^T u \rangle \nonumber\\
&= u^T \K{(}{)}{\Sigma P^T \Gamma_K + \partial_K (\Sigma P^T)} \nabla_{P\Sigma^T u} | \Lambda^T P \Sigma^T u \rangle.\label{canonreduced}
\end{align}
We now introduce the differential $\tilde{\Sigma}$ from $y$ to $z$ coordinates. In particular:
\begin{equation*}
\tilde{\Sigma}_{\mu}^{\alpha} := \frac{\partial y^{\alpha}}{\partial z^{\mu}}= \lambda^{-1}\K{(}{)}{\delta_{\mu}^{\alpha} - y^{\alpha}\delta_{\mu}^0}.
\end{equation*}
Then we have:
\begin{equation*}
P^T \tilde{\Sigma} \Sigma P^T = P^T.
\end{equation*}
Therefore:
\begin{equation*}
\nabla_{P\Sigma^T u} = P^T \tilde{\Sigma} \nabla_u + (1 - P^T) \nabla_{P\Sigma^T u}.
\end{equation*}
Substituting in (\ref{canonreduced}) we obtain:
\begin{align*}
&\partial_K |\Lambda^T P\Sigma^T u \rangle=\\
&u^T \K{(}{)}{\tilde{\Gamma}_K \nabla_u +  \K{(}{)}{\Sigma P^T \Gamma_K + \partial_K (\Sigma P^T)} (1 - P^T)\nabla_{P\Sigma^T u}}|\Lambda^T P\Sigma^T u \rangle.
\end{align*}
In the above we have defined the connection:
\begin{equation}
\tilde{\Gamma}_K = \Sigma P^T \Gamma_K P^T \tilde{\Sigma} + \partial_K(\Sigma P^T) P^T \tilde{\Sigma}.\label{tildeGamma}
\end{equation}
As we show in appendix \ref{appB}, the connection $\tilde{\Gamma}$ splits into purely holomorphic and anti-holomorphic components with $\tilde{\Gamma}_{\overline{\alpha} \overline{\beta}}^{\overline{\gamma}} = (\tilde{\Gamma}_{\alpha \beta}^{\gamma})^*$ and can be expressed in terms of the Levi-Civita connection $\hat{\Gamma}$ of $\hat{g}$ as follows:
\begin{align*}
\tilde{\Gamma}_{\alpha \beta}^{\gamma} &= \hat{\Gamma}_{\alpha \beta}^{\gamma} + \partial_{\alpha} \log |K| \delta_{\beta}^{\gamma}\\
\tilde{\Gamma}_{0 \beta}^{\gamma} &= \lambda^{-1} \delta_{\beta}^{\gamma},
\end{align*}
where we have denoted by $0$ the coordinate $\lambda$. In particular $\tilde{\Gamma}$ is compatible with the metric $\tilde{g}$.

\medskip
\noindent
Now we turn to the dependence on $p$ of the coherent state $\mathcal{P}(p, A)|\Sigma^T u \rangle_{p, A}$ proper. In fact, to tackle the reduced tensor $\mathcal{C}_{red}$ directly we compute:
\begin{align*}
&\partial_K \K{(}{)}{\exp \K{(}{)}{\frac{1}{4}||P(p)\Sigma^T u||^2_{g(p)}} | P\Sigma^T u \rangle_{p, A}}
=\\
&\K{(}{)}{ - i\beta_K - \frac{i}{2}(\Gamma_K \omega)_{IJ} \hat{x}^I \hat{x}^J + u^r\tilde{\Gamma}_{Kr}^s \frac{\partial}{\partial u^s}+ u^r \K{(}{)}{\K{(}{)}{\Sigma P^T \Gamma_K + \partial_K (\Sigma P^T)} (1 - P^T)}_r^S \frac{\partial}{\partial (P\Sigma^Tu)^S}}\cdot\\
&\exp \K{(}{)}{\frac{1}{4}||P(p)\Sigma^T u||^2_{g(p)}} | P\Sigma^T u \rangle_{p, A}.
\end{align*}
In the above we have used the metric compatibility of $\tilde{\Gamma}$. At this point we need to compute the action of $\hat{x}$ on the coherent state. However only the components of $\hat{x}$ along the horizontal bundle act naturally as differential operators. We shall now focus our attention on those:
\begin{align*}
&\tilde{\Sigma}^T P\hat{x} \,  \K{(}{)}{\exp \K{(}{)}{\frac{1}{4}||P(p)\Sigma^T u||^2_{g(p)}}| P \Sigma^T u \rangle_{p, A}}\\
&= \K{(}{)}{\frac{1}{2} \tilde{\Sigma}^T P(1 - iJ)^T P\Sigma^T u + \frac{1}{2} \tilde{\Sigma}^T P(g^{-1} - i \omega^{-1}) P^T \tilde{\Sigma}\nabla_u}  \exp \K{(}{)}{\frac{1}{4}||P(p)\Sigma^T u||^2_{g(p)}}| P\Sigma^T u \rangle_{p, A}\\
&= \K{(}{)}{\frac{1}{2}(1 - i\hat{J})^T u - \frac{1}{2K} (\hat{g}^{-1} - i \hat{\omega}^{-1})\nabla_u}  \exp \K{(}{)}{-\frac{1}{4}K ||u||^2_{\hat{g}(p)}}| P\Sigma^T u \rangle_{p, A},
\end{align*}
Now we shall present the explicit form of the master equation:
\begin{equation*}
\Sigma P^T \! _{p, A}\langle \Sigma^T u | \mathcal{P}^{\dagger}(p, A) (-1)^F (d + A) | p \rangle_A = 0.
\end{equation*}
Using (\ref{defup2}), we arrive at the master equation for $\mathcal{C}_{red}(u^i \partial_i)$:
\begin{align}
&\K{(}{.}{(\Sigma P^T \nabla_x)_k - u^r (\hat{\Gamma}_k)_r^s \frac{\partial}{\partial u^s} + i (\Sigma P^T(\alpha - \beta))_k}\nonumber\\
&+ iK\hat{\omega}_{ik}  \K{(}{)}{\frac{1}{2}(1 - i\hat{J})^T u - \frac{1}{2K} (\hat{g}^{-1} - i \hat{\omega}^{-1})\nabla_u}^i\nonumber\\
&+ \frac{i}{2}C_{kij} \K{(}{)}{\frac{1}{2K} (\hat{g}^{-1} - i \hat{\omega}^{-1})\nabla_u}^i
\K{(}{)}{\frac{1}{2K} (\hat{g}^{-1} - i \hat{\omega}^{-1})\nabla_u}^j\nonumber\\
&+ \frac{i}{2}C_{kij} \K{(}{)}{\frac{1}{2}(1 - i\hat{J})^Tu}^i\K{(}{)}{\frac{1}{2}(1 - i\hat{J})^Tu}^j \nonumber\\
&- \K{.}{)}{u^r (\Sigma P^T)_k^K\K{(}{)}{ \K{(}{)}{\Sigma P^T \Gamma_K + \partial_K (\Sigma P^T)} (1 - P^T)}_r^S \frac{\partial}{\partial (P\Sigma^Tu)^S}}\mathcal{C}_{red}(u^i \partial_i) = 0. \label{fibreterm}
\end{align}
Before expressing the master equation in holomorphic and anti-holomorphic parts, we shall decompose the last term of (\ref{fibreterm}) in holomorphic and anti-holomoprhic parts. Since $\mathcal{C}_{red}$ has only anti-holomorphic legs, in complex coordinates the holomorphic part is given by:
\begin{align*}
&-(\Sigma P^T)_{\beta}^{\sigma}dy^{\overline{\alpha}} \Sigma_{\overline{\alpha}}^{\overline{\mu}} (\partial_{\sigma} P_{\overline{\mu}}^{\overline{\nu}}) (1 - P)_{\overline{\nu}}^{\overline{\rho}} \iota_{\partial_{\overline{\rho}}} \mathcal{C}_{red}\\
&= (\Sigma P^T)_{\beta}^{\sigma}dy^{\overline{\alpha}} \partial_{\sigma} \partial_{\overline{\alpha}}\log |K| z^{\overline{\rho}}\iota_{\partial_{\overline{\rho}}}\mathcal{C}_{red}\\
&= dy^{\overline{\alpha}} \hat{g}_{\beta \overline{\alpha}} z^{\overline{\rho}} \iota_{\partial_{\overline{\rho}}} \mathcal{C}_{red},
\end{align*}
while the anti-holomorphic part reads:
\begin{align*}
&-(\Sigma P^T)_{\overline{\beta}}^{\overline{\sigma}}dy^{\overline{\alpha}} \Sigma_{\overline{\alpha}}^{\overline{\mu}} (P_{\overline{\mu}}^{\overline{\tau}} \Gamma_{\overline{\sigma} \overline{\tau}}^{\overline{\nu}} + \partial_{\overline{\sigma}} P_{\overline{\mu}}^{\overline{\nu}}) (1 - P)_{\overline{\nu}}^{\overline{\rho}} \iota_{\partial_{\overline{\rho}}} \mathcal{C}_{red}\\
&=  0.
\end{align*}
The above is a result of the following identity:
\begin{align*}
&(P_{\overline{\mu}}^{\overline{\tau}} \Gamma_{\overline{\sigma} \overline{\tau}}^{\overline{\nu}} + \partial_{\overline{\sigma}} P_{\overline{\mu}}^{\overline{\nu}}) (1 - P)_{\overline{\nu}}^{\overline{\rho}}\\
&= (\Gamma_{\overline{\sigma} \overline{\mu}}^{\overline{\nu}} + \partial_{\overline{\sigma}}P_{\overline{\mu}}^{\overline{\nu}})(1 - P)_{\overline{\nu}}^{\overline{\rho}}\\
&= (g^{\overline{\nu} \nu}\partial_{\overline{\sigma}}g_{\nu \overline{\mu}} + \partial_{\overline{\sigma}}P_{\overline{\mu}}^{\overline{\nu}})(1 - P)_{\overline{\nu}}^{\overline{\rho}}\\
&= \frac{1}{K}\partial_{\overline{\sigma}}(z^{\nu} g_{\nu \overline{\mu}})z^{\overline{\rho}} - (\delta_{\overline{\sigma}}^{\overline{\nu}} \partial_{\overline{\mu}} \log |K| + z^{\overline{\nu}} \partial_{\overline{\sigma}}\partial_{\overline{\mu}} \log|K|)z^{\overline{\rho}}\partial_{\overline{\nu}}\log|K| \\
&= \frac{\partial_{\overline{\sigma}}\partial_{\overline{\mu}}K}{K} z^{\overline{\rho}} - \frac{\partial_{\overline{\sigma}}\partial_{\overline{\mu}}K}{K} z^{\overline{\rho}}\\
&= 0.
\end{align*}
In the first step above we have used the fact that in special coordinates:
\begin{align}
&z^{\overline{\rho}}\Gamma_{\overline{\rho}} = g^{-1}z^{\overline{\rho}}\partial_{\overline{\rho}}g = 0.\label{GammaH}
\end{align}
We are thus left to compute
\begin{equation*}
z^{\overline{\rho}}\iota_{\partial_{\overline{\rho}}} \mathcal{C}_{red}. 
\end{equation*}
For this we need to resort to (\ref{masterantiholo1}) using (\ref{GammaH}). We obtain:
\begin{align*}
z^{\overline{\rho}}\iota_{\partial_{\overline{\rho}}}\mathcal{C}_{red} &= \K{(}{)}{\iota^*\K{(}{)}{z^{\overline{\rho}}\iota_{\partial_{\overline{\rho}}}\mathcal{C}} \circ P}\\
&= \K{(}{)}{\iota^*\K{(}{)}{\K{(}{)}{-z^{\overline{\rho}} \partial_{\overline{\rho}} - i (\alpha - \beta)(\overline{H})}\mathcal{C}} \circ P}\\
&= \K{(}{)}{- \overline{\lambda} \frac{\partial}{\partial \overline{\lambda}} + dy^{\overline{\alpha}}\iota_{\partial_{\overline{\alpha}}} - i(\alpha - \beta)\K{(}{)}{\overline{\lambda} \frac{\partial}{\partial \overline{\lambda}}}}\mathcal{C}_{red}.
\end{align*}

\medskip
\noindent
In order to isolate the dependence of $\mathcal{C}_{red}$ on $\lambda$ and $\overline{\lambda}$, we use the fact that in $y$ coordinates $C_{\alpha \beta \gamma}$ is holomorphic homogeneous of degree $2$ in $\lambda$. Thus, we define the normalized $C$ tensor through:
\begin{equation*}
C_{\alpha \beta \gamma}(\lambda, y) = \lambda^2 \hat{C}_{\alpha \beta \gamma}(y).
\end{equation*}

\medskip
\noindent
At this point we have all the ingredients to express the master equation (\ref{fibreterm}) in holomorphic and anti-holomoprhic parts. The anti-holomorphic part reads:
\begin{align}
&\K{(}{.}{\nabla^{(0,1)}_{\overline{\alpha}} - (\partial_{\overline{\alpha}} \log h) \overline{\lambda} \frac{\partial}{\partial \overline{\lambda}} + i(\alpha - \beta)_{\overline{\alpha}} - i(\partial_{\overline{\alpha}} \log h) (\alpha - \beta)\K{(}{)}{\overline{\lambda}\frac{\partial}{\partial \overline{\lambda}}}}\nonumber\\
&+\K{.}{)}{\iota_{\partial_{\overline{\alpha}}} + \frac{i \overline{\lambda}^2}{2} \hat{C}_{\overline{\alpha}\overline{\beta}\overline{\gamma}} dy^{\overline{\beta}}dy^{\overline{\gamma}}
}\mathcal{C}_{red} = 0,\label{projmaster1}
\end{align}
while the holomorphic part reads:
\begin{align*}
&\K{(}{.}{\partial_{\alpha} - (\partial_{\alpha} \log h) \lambda \frac{\partial}{\partial \lambda} + i(\alpha - \beta)_{\alpha} - i(\partial_{\alpha} \log h) (\alpha - \beta)\K{(}{)}{\lambda\frac{\partial}{\partial \lambda}}}\\
& + dy^{\overline{\beta}} \hat{g}_{\alpha \overline{\beta}} \K{(}{)}{- \overline{\lambda} \frac{\partial}{\partial \overline{\lambda}} + dy^{\overline{\gamma}}\iota_{\partial_{\overline{\gamma}}} - i(\alpha - \beta)\K{(}{)}{\overline{\lambda} \frac{\partial}{\partial \overline{\lambda}}}} \\
&- \K{.}{)}{iK \hat{\omega}_{\alpha \overline{\beta}} dy^{\overline{\beta}} + \frac{i\lambda^2}{2K^2} \hat{C}_{\alpha \beta \gamma} \hat{g}^{\beta \overline{\beta}}\hat{g}^{\gamma \overline{\gamma}} \iota_{\partial_{\overline{\beta}}}\iota_{\partial_{\overline{\gamma}}}}\mathcal{C}_{red} = 0.
\end{align*}
There are of course two further equations left, inherited from the master equation of the conic affine special K\"ahler manifold $M$. We have already made full use of the anti-holomorphic part to express the last term in (\ref{fibreterm}) as a differential operator on $\mathcal{C}_{red}$. From the holomorphic part we obtain instead:
\begin{equation}
\K{(}{)}{\frac{\partial}{\partial \lambda} + i(\alpha - \beta)\K{(}{)}{\frac{\partial}{\partial \lambda}}} \mathcal{C}_{red}(u^i\partial_i) = 0,\label{fibreholo}
\end{equation}
which in components reads:
\begin{align*}
\K{(}{)}{\frac{\partial}{\partial \lambda} + i(\alpha - \beta) \K{(}{)}{\frac{\partial}{\partial \lambda}}} \mathcal{C}_{red}^0 &= 0\\
\mathcal{C}_{red}^n (\lambda, \overline{\lambda}, y, \overline{y}) &= \tilde{\mathcal{C}}_{red}^n(\overline{\lambda}, y, \overline{y}) \s \forall n \geq 1.
\end{align*}
With (\ref{fibreholo}) the holomorphic part of the master equation simplifies to:
\begin{align}
&\K{(}{.}{\partial_{\alpha} + i(\alpha - \beta)_{\alpha} + dy^{\overline{\beta}} \hat{g}_{\alpha \overline{\beta}} \K{(}{)}{- \overline{\lambda} \frac{\partial}{\partial \overline{\lambda}} + dy^{\overline{\gamma}}\iota_{\partial_{\overline{\gamma}}}}  - dy^{\overline{\beta}} \hat{g}_{\alpha \overline{\beta}}\K{(}{)}{i(\alpha - \beta)\K{(}{)}{\overline{\lambda} \frac{\partial}{\partial \overline{\lambda}}} - K}}\nonumber\\
&+\K{.}{)}{ \frac{i\lambda^2}{2K^2} \hat{C}_{\alpha \beta \gamma} \hat{g}^{\beta \overline{\beta}}\hat{g}^{\gamma \overline{\gamma}} \iota_{\partial_{\overline{\beta}}}\iota_{\partial_{\overline{\gamma}}}}\mathcal{C}_{red} = 0.\label{projmaster2}
\end{align}
As a last step we will choose for $\alpha$ and $\beta$ the gauge adopted in section \ref{MEsection} and we will express the master equation (\ref{projmaster1}, \ref{fibreholo}, \ref{projmaster2}) as an equation for $\mathcal{S}_{red}$, which analogously to $\mathcal{S}$ in section \ref{MEsection}, is defined as:
\begin{equation*}
\mathcal{S}_{red} = \K{(}{)}{\mathrm{det}\,g}^{\frac{1}{8}}e^{\frac{K}{2}}\mathcal{C}_{red}.
\end{equation*}
Noticing that in this gauge:
\begin{equation*}
\beta\K{(}{)}{H} = \beta\K{(}{)}{\overline{H}} = 0,
\end{equation*}
we obtain that (\ref{projmaster1}, \ref{projmaster2}):
\begin{align}
&\K{(}{)}{\nabla^{(0,1)}_{\overline{\alpha}} - (\partial_{\overline{\alpha}}\log h) \overline{\lambda}\frac{\partial}{\partial \overline{\lambda}}- 2i\beta_{\overline{\alpha}} +\iota_{\partial_{\overline{\alpha}}} + \frac{i \overline{\lambda}^2}{2} \hat{C}_{\overline{\alpha}\overline{\beta}\overline{\gamma}} dy^{\overline{\beta}}dy^{\overline{\gamma}}
}\mathcal{S}_{red} = 0,\label{projmasterS1}\\
&\K{(}{)}{\partial_{\alpha} + dy^{\overline{\beta}} \hat{g}_{\alpha \overline{\beta}} \K{(}{)}{- \overline{\lambda} \frac{\partial}{\partial \overline{\lambda}} + dy^{\overline{\gamma}}\iota_{\partial_{\overline{\gamma}}}} +\frac{i\lambda^2}{2K^2} \hat{C}_{\alpha \beta \gamma} \hat{g}^{\beta \overline{\beta}}\hat{g}^{\gamma \overline{\gamma}} \iota_{\partial_{\overline{\beta}}}\iota_{\partial_{\overline{\gamma}}}}\mathcal{S}_{red} = 0,\label{projmasterS2}
\end{align}
while (\ref{fibreholo}) becomes:
\begin{align}
\frac{\partial}{\partial \lambda} \mathcal{S}_{red} = 0.\label{fibreholoS}
\end{align}
We have finally arrived at the precise generalization (\ref{projmasterS1}, \ref{projmasterS2}, \ref{fibreholoS}) of the holomorphic anomaly equation of \cite{bcov9309140} while at the same time having provided its general solution (\ref{lorentzsol}, \ref{Cred}).

\newpage

\section{Concluding remarks}
In the present paper we have shown how special K\"ahler manifolds arise from the structure of quantization, and constructed their quantum counterpart. Crucial to our constructions was the central idea developed in \cite{witten} and the formalism of \cite{fedosov94}. We have shown how a general version of the holomorphic anomaly equation of \cite{bcov9309140} arises in our construction while at the same time providing its general solution. 

\medskip
\noindent
The present work needs however to be further developed to understand better the physical, string theoretic, meaning of these solutions. In particular it is still to be understood, from a quantization perspective, how to isolate the analogue of the generating function of closed topological strings in a given D-brane configuration \cite{neitzke-walcher} . In this regard, it seems as though a starting point for these developments within this work could be the discussion at the end of section \ref{quantumconic}.

\medskip
\noindent
\subsection*{Acknowledgements}

I would like to thank Ilka Brunner for her constant support and encouragement during the course of this work. I am also grateful to her, Patrick B\"ohl, Nils Carqueville and Andr\'es Collinucci for useful comments on the manuscript.

\newpage

\appendix

\section{A few identities of Special Geometry}\label{appendix}
Here we shall just give the form of the Ricci tensor for an affine special K\"ahler manifold, as it is needed in section \ref{MEsection}. For the sake of coherence we will compute it in special Darboux coordinates. We shall need the expression for the Christoffel symbols, that reduces to
\begin{equation*}
\Gamma_{ij}^k = \frac{1}{2}g^{kr}\partial_r\partial_i\partial_j K
\end{equation*}
and in particular the following identity:
\begin{align*}
J_i^r \partial_r \partial_k \partial_l K &= -(\partial_k J_i^r)g_{rl}\\
&= -\K{(}{)}{\partial_k(g_{is}\omega^{sr})}g_{rl}\\
&= \partial_k\partial_i\partial_s K J_l^s  \s.
\end{align*}
Equivalently the tensor $C$, which in special Darboux coordinates reads
\begin{equation*}
C_{ijk} = \frac{1}{2}J_i^r\partial_r\partial_j\partial_kK \s,
\end{equation*}
is symmetric. Moreover the fact that $\partial_i\partial_j\partial_k K$ is symmetric implies that $C$ splits into holomorphic and anti-holomorphic parts. From the above, in particular, it follows:
\begin{align*}
\Gamma_{ki}^k &= \frac{1}{2}g^{kr}\partial_r\partial_k\partial_iK\\
&= -\frac{1}{2}\omega^{ks}J_s^r\partial_r\partial_k\partial_iK\\
&= -\frac{1}{2}\omega^{ks}\partial_r\partial_k\partial_s J_i^r\\
&= 0,
\end{align*}
hence:
\begin{align*}
R_{ij} &= \partial_k \Gamma_{ij}^k - \partial_j \Gamma_{ki}^k + \Gamma_{kl}^k \Gamma_{ji}^l - \Gamma_{jl}^k \Gamma_{ki}^l\\
&=  \partial_k \Gamma_{ij}^k - \Gamma_{jl}^k \Gamma_{ki}^l \s.
\end{align*}
The first term can be rewritten as follows:
\begin{align*}
\partial_k \Gamma_{ij}^k &= \frac{1}{2}(\partial_k g^{kr})\partial_i\partial_j\partial_r K + \frac{1}{2}g^{kr} \partial_k\partial_i\partial_j\partial_rK\\
&= \frac{1}{2}g^{kr} \partial_k\partial_i\partial_j\partial_rK\\
&= -\frac{1}{2}\partial_i g^{kr} \partial_k\partial_j \partial_r K\\
&= \frac{1}{2}\omega^{ks}\partial_i\partial_s\partial_l K \omega^{lr} \partial_k \partial_j \partial_r K\\
&= 2 g^{ks}C_{isl} g^{lr}C_{kjr} \s.
\end{align*}
The second term in the expression for the Ricci tensor can instead be rewritten as:
\begin{align*}
-\Gamma_{jl}^k\Gamma_{ki}^l &= -\frac{1}{4}g^{kr} \partial_r\partial_j\partial_l K g^{ls}\partial_s\partial_k\partial_i K\\
&= \frac{1}{4} g^{kr}J_r^t J_t^u \partial_u \partial_j \partial_lK g^{ls} \partial_s \partial_k \partial_i K\\
&= \frac{1}{4} g^{kr}J_r^t \partial_t\partial_j\partial_u K J_l^u g^{ls} \partial_s \partial_k \partial_iK\\
&= -\frac{1}{4}g^{kr}J_r^t \partial_t\partial_j\partial_u K g^{lu} J_l^s \partial_s \partial_k \partial_iK\\
&= -g^{kr}C_{rju}g^{lu}C_{lki} \s.
\end{align*}
Thus, finally:
\begin{equation*}
R_{ij} = g^{ks}C_{isl} g^{lr}C_{kjr}.
\end{equation*}

\section{The connection on the horizontal bundle}\label{appB}
Here we analyze the connection $\tilde{\Gamma}$ defined in (\ref{tildeGamma}) and express it in terms of the Levi-Civita connection $\hat{\Gamma}$ of $\hat{g}$.

\medskip
\noindent
First we will show that $\tilde{\Gamma}$ is compatible with the metric $\tilde{g}$. We thus compute:
\begin{align*}
&\K{(}{)}{\partial_K - u^T \tilde{\Gamma}_K \nabla_u}||u||_{\tilde{g}}^2\\
&=u^T\partial_K\K{(}{)}{\Sigma P^T g P \Sigma^T}u - 2u^T \K{(}{)}{(\Sigma P^T \Gamma_K P^T \tilde{\Sigma} + \partial_K(\Sigma P^T) P^T \tilde{\Sigma})\Sigma P^T g P \Sigma^T}u\\
&= u^T\partial_K\K{(}{)}{\Sigma P^T g P \Sigma^T}u - 2u^T \K{(}{)}{\Sigma P^T \Gamma_K P^T g\Sigma^T + \partial_K(\Sigma P^T) P^T g P \Sigma^T}u\\ 
&= u^T\partial_K\K{(}{)}{\Sigma P^T g P \Sigma^T}u - 2u^T \K{(}{)}{\Sigma P^T \Gamma_K g\Sigma^T + \partial_K(\Sigma P^T) g P \Sigma^T}u\\
&= 0.
\end{align*}

\medskip
\noindent
Now we shall express $\tilde{\Gamma}$ in terms of $\hat{\Gamma}$. We start by expressing the latter using the fact that $\hat{g}$ is K\"ahler:
\begin{align*}
\hat{\Gamma}_{\alpha \beta}^{\gamma} =& \K{(}{)}{K \tilde{\Sigma}^T P g^{-1} P^T \tilde{\Sigma} \partial_{\alpha}\K{(}{)}{\frac{1}{K} \Sigma P^T g P \Sigma^T}}_{\beta}^{\gamma}\\
=&\K{(}{)}{\tilde{\Sigma}^T P g^{-1} \overline{P}^T \overline{\tilde{\Sigma}} \K{(}{)}{\partial_{\alpha}(\overline{\Sigma}\overline{P}^T) g P\Sigma^T + \overline{\Sigma} \overline{P}^T\partial_{\alpha} g P\Sigma^T + \overline{\Sigma} \overline{P}^T g \partial_{\alpha}(P\Sigma^T)}}_{\beta}^{\gamma}\\
&-\partial_{\alpha} \log |K| \delta_{\beta}^{\gamma}\\
=& \tilde{\Sigma}^T P g^{-1} \overline{P}^T \overline{\tilde{\Sigma}} \overline{\Sigma}\K{(}{)}{\partial_{\alpha}\overline{P}^T g P\Sigma^T} + \tilde{\Gamma}_{\alpha \beta}^{\gamma} - \partial_{\alpha} \log |K| \delta_{\beta}^{\gamma}\\
=& \tilde{\Gamma}_{\alpha \beta}^{\gamma} - \partial_{\alpha} \log |K| \delta_{\beta}^{\gamma}.
\end{align*}
The remaining components of $\tilde{\Gamma}$ are given by:
\begin{align*}
\tilde{\Gamma}_{\overline{\alpha} \beta}^{\gamma} &= \K{(}{)}{\partial_{\overline{\alpha}}(\Sigma P^T) P^T \tilde{\Sigma}}_{\beta}^{\gamma}\\
&= \Sigma_{\beta}^{\mu}\partial_{\overline{\alpha}}P_{\mu}^{\nu}\lambda^{-1}(\delta_{\nu}^{\gamma} - y^{\gamma}\delta_{\nu}^0)\\
&= \Sigma_{\beta}^{\mu}\partial_{\overline{\alpha}}\partial_{\mu} \log |K| \lambda^{-1}z^{\nu}(\delta_{\nu}^{\gamma} - y^{\gamma}\delta_{\nu}^0)\\
&= 0.
\end{align*}
Similarly:
\begin{align*}
\tilde{\Gamma}_{\overline{0}\beta}^{\gamma} = 0,
\end{align*}
while
\begin{align*}
\tilde{\Gamma}_{0\beta}^{\gamma} &= \K{(}{)}{\frac{\partial}{\partial \lambda}(\Sigma P^T)P^T \tilde{\Sigma}}_{\beta}^{\gamma}\\
&= \frac{\partial}{\partial \lambda} \K{(}{)}{\lambda \delta_{\beta}^{\mu} - z^{\mu}\partial_{\beta}\log |K|}\lambda^{-1}(\delta_{\mu}^{\gamma} - y^{\gamma}\delta_{\mu}^0)\\
&= \lambda^{-1}\delta_{\beta}^{\gamma}.
\end{align*}
Finally:
\begin{align*}
\tilde{\Gamma}_{\overline{\alpha} \overline{\beta}}^{\overline{\gamma}} &= (\tilde{\Gamma}_{\alpha \beta}^{\gamma})^*\\
\tilde{\Gamma}_{\overline{0}\overline{\beta}}^{\overline{\gamma}} &= (\tilde{\Gamma}_{0\beta}^{\gamma})^*.
\end{align*}

\end{document}